\documentclass[12pt, draftclsnofoot, onecolumn]{IEEEtran}
\usepackage{amsmath,epsfig,amssymb,verbatim,amsopn,subfigure,color}
\usepackage{cite,xspace}
\usepackage{array,algorithm,algorithmic}
\usepackage{bbm}
\usepackage{array}
\usepackage{multirow}
\usepackage{multicol}
\usepackage{rotating}
\usepackage{dsfont,float}
\usepackage{stfloats}
\usepackage{fancyhdr}
\usepackage{graphicx}
\usepackage{enumerate,makecell}
\usepackage{xcolor}
\usepackage{datetime}
\usepackage{diffcoeff}
\allowdisplaybreaks

\normalsize

\ifCLASSOPTIONpeerreview

\fi

\ifCLASSOPTIONonecolumn
    
\else
    
\fi

\makeatletter
\let\l@ENGLISH\l@english
\makeatother

\makeatletter
\renewcommand*{\@opargbegintheorem}[3]{\trivlist
  \item[\hskip \labelsep{\itshape #1\ #2}] {\itshape (#3):} {\normalfont}}
\makeatother
\usepackage{relsize}

\newcommand{\cellrad} {R_{U}}

\newcommand{\hovtime} {T_{\mathrm{hov}}}

\newcommand{\slottime} {t_{\mathrm{s}}}
\newcommand{\frametime} {t_{\mathrm{f}}}
\newcommand{\nslot} {n_{\mathrm{s}}}

\newcommand{\hfad} {h}

\newcommand{\hUAV} {H_{\mathrm{U}}}

\newcommand{\pathlossIOT} {\zeta_{1}}

\newcommand{\powbud} {P_{\mathrm{max}}}

\newcommand{\Psucc}{P_{\mathrm{sc}}}
\newcommand{\NAusers}{N_{\mathrm{A}}}
\newcommand{\databits}{D}

\newcommand{\bandwidth}{B}
\newcommand{\errorProb}{\epsilon}
\newcommand{\nAVGpack}{\lambda}
\newcommand{\threshold}{\theta_{\mathrm{T}}}
\newcommand{\avgTransP}{\bar{\mathrm{P}}}
\newcommand{\noiseP}{\sigma^{2}}

\newcommand{\PLcoef}{\eta}
\newcommand{\PLindex}{\alpha}
\newcommand{\Intsing}{I_{\mathrm{s}}}
\newcommand{\Intcol}{I_{\mathrm{c}}}
\newcommand{\Pmax}{P_{\mathrm{max}}}
\newcommand{\densing}{\varphi_{s}}
\newcommand{\dencol}{\varphi_{c}}
\newcommand{\PPPsing}{\phi_{s}}
\newcommand{\PPPcol}{\phi_{c}}
\newcommand{\PPPori}{\phi}
\newcommand{\denori}{\varphi}
\newcommand{\Pcf}{P_{\mathrm{cf}}}
\newcommand{\Plambda}{P_{\lambda}}
\newcommand{\Nmusa}{N_{\mu}}
\newcommand{\Nsing}{L_{s}}
\newcommand{\PAct}{P_{A}}
\newcommand{\NAslot}{L_{A}}
\newcommand{\imgI}{\mathrm{i}}
\newcommand{\barnes}{\mathrm{G}}
\newcommand{\harmonic}{\mathrm{H}}
\newcommand{\SINR}{\gamma}

\newtheorem{lemma}{Lemma}
\newtheorem{remark}{Remark}
\newtheorem{theorem}{Theorem}
\newtheorem{proposition}{Proposition}

\newtheorem{assumption}{Assumption}

\newtheorem{definition}{Definition}
\graphicspath{{figs/},{Figures/}}

\newcommand{\AuthorOne}{Nilupuli~Senadhira, {\em{Student Member, IEEE}}}
\newcommand{\AuthorTwo}{Salman~Durrani, {\em{Senior Member, IEEE}}}
\newcommand{\AuthorThree}{Sheeraz~A.~Alvi, {\em{Member, IEEE}}}
\newcommand{\AuthorFour}{Nan~Yang, {\em{Senior Member, IEEE}}}
\newcommand{\AuthorFive}{Xiangyun~Zhou, {\em{Fellow, IEEE}}}

\newcommand{\ThankOne}{N. Senadhira, S. Durrani, S. Alvi, N. Yang and X. Zhou are with the School of Engineering, The Australian National University, Canberra, Australia (Corresponding author email:  salman.durrani@anu.edu.au).}

\newcommand{\ThankTwo}{This work \textcolor{black}{was} presented in part at the IEEE ICC, Rome, Italy, May 2023~\cite{Nilu-2023}.}

\begin{document}

\title{UAV-assisted IoT Monitoring Network: Adaptive Multiuser Access for Low-Latency and High-Reliability Under Bursty Traffic}
\author{\IEEEauthorblockN{\AuthorOne,~\AuthorTwo,~\AuthorThree,~\AuthorFour, and \AuthorFive}\thanks{\ThankOne}\thanks{\ThankTwo}} 
\maketitle \vspace{-5mm}

\begin{abstract}
In this work, we propose an adaptive system design for an Internet of Things (IoT) monitoring network with latency and reliability requirements, where IoT devices generate time-critical and event-triggered bursty traffic, and an unmanned aerial vehicle (UAV) aggregates and relays sensed data to the base station. Existing transmission schemes based on the overall average traffic rates over-utilize network resources when traffic is smooth, and suffer from packet collisions when traffic is bursty which occurs in an event of interest. We address such problems by designing an adaptive transmission scheme employing multiuser shared access (MUSA) based grant-free non-orthogonal multiple access and use short packet communication for low latency of the IoT-to-UAV communication. Specifically, to accommodate bursty traffic, we design an analytical framework and formulate an optimization problem to maximize the performance by determining the optimal number of transmission time slots, subject to the stringent reliability and latency constraints. We compare the performance of the proposed scheme with a non-adaptive power-diversity based scheme with a fixed number of time slots. Our results show that the proposed scheme has superior reliability and stability in comparison to the state-of-the-art scheme at moderate to high average traffic rates, while satisfying the stringent latency requirements.
\end{abstract}


\newpage
\section{Introduction}
According to the International Disasters Database, 432 catastrophic events were recorded globally in 2021, which is significantly higher than the average of 357 annual extreme weather events recorded between 2001 and 2020~\cite{EMDAT}. Such events come at a huge ecological, social and financial cost. For instance, it is estimated that the direct cost from extreme weather events will reach $\$35$ billion by 2050 in Australia \cite{lefebvre2022cost}. Thus, there is a great interest in IoT-based environmental monitoring systems which can provide early warnings to help mitigate economic and social impacts and to improve the disaster resilience of communities and countries.


Recently, IoT-based environmental monitoring networks \textcolor{black}{(EMNs)} have been deployed to monitor an event of interest (EoI) in practical scenarios such as (i) bushfire detection \cite{yebra2021integrated}, (ii) hazardous gas detection in a volcanic area \cite{awadallah2019an}, and (iii) water pollution monitoring systems \cite{neha2016smart}. These systems generate heterogenous traffic, depending on environmental conditions. Under the normal condition, IoT devices measure environmental data and send smooth, non-real time, and low-rate traffic data to a control center \cite{nikaein2013simple,bulashenko2020new}. However, when an EoI is detected, IoT devices generate uplink dominant and time-critical high-rate traffic, which may include information about the crisis event and corrective actions to resolve the event. \textit{Thus, it is important to design IoT-based \textcolor{black}{EMNs} to meet high-reliability and low-latency requirements when an EoI occurs, which motivates this work.}.

\underline{\textit{UAV for IoT-based \textcolor{black}{EMNs}:}}
The IoT devices are typically low-cost battery-limited sensors with limited computational capabilities, and have short transmission range which is inadequate to reach high altitude platforms and satellites \cite{bushnaq2019aeronautical}. If the IoT devices are deployed in an area with poor wireless coverage, e.g., in a mountain or forest area with no terrestrial cellular infrastructure, the reliable and low-latency transmission of data becomes a major challenge \cite{dai2020efficient}. In such deployment, unmanned aerial vehicles (UAVs) can be strategically deployed in IoT environments to increase the communication range of the IoT devices and improve the reliability of IoT communications \cite{mozaffari2017mobile,sun2021two,
wu2021comprehensive}. Consequently, the IoT devices can transmit data with low transmit power, resulting in prolonged battery life. Also, due to the aerial nature, UAVs have the capability to establish line-of-sight communication links with the IoT devices. This would further favor IoT communications by mitigating shadowing and signal blockage in communication links compared to terrestrial communications \cite{dai2020efficient,mozaffari2017mobile,lei2022performance}. Thus, UAVs are considered as an attractive solution to IoT data aggregation before transmitting to a cellular base station (BS).

\underline{\textit{Multiple Access Protocol for IoT-based \textcolor{black}{EMNs}:}}
To support the connectivity of IoT devices, two main access protocols have been proposed in the literature \cite{shahab2020grant}, namely, orthogonal multiple access (OMA) and non-orthogonal multiple access (NOMA). In OMA, the radio resources are only occupied by a single device.  In NOMA, this limitation is overcome by multiplexing devices over a single channel resource \cite{di2017noma,zhang2022semi}. Thus, NOMA is suitable for IoT-based \textcolor{black}{EMNs} due to its large availability for using channel resources.

The channel resource access protocol of multiplexing users in NOMA needs to be carefully chosen. In the conventional grant-based (GB) access, the users perform uplink scheduling requests via contention-based random access, which is known as a performance bottleneck due to its high latency and signaling overhead \cite{shahab2020grant}. On the other hand, grant-free (GF) access, where the IoT devices are able to wake up and transmit, is preferred for low-latency transmission and also reduces the power consumption of the IoT devices \cite{yuan2016multi}. Thus, GF-NOMA is a suitable multiple-access protocol for IoT-based \textcolor{black}{EMNs}.

Recently, multiuser shared access (MUSA) has been proposed as a signature-based GF-NOMA scheme \cite{shahab2020grant}. MUSA utilizes smart interfering management methods, i.e., successive interference cancellation (SIC), and exploits device-specific signatures to decode the data at the receiver \cite{shahab2020grant}. In MUSA, devices autonomously and randomly choose complex-valued short spreading codes with low cross-correlation whenever they have data to transmit. Thus, no resource coordination is required at the devices. Due to the additional design freedom provided by the real and imaginary components of the spreading code, the active devices have the option to choose from a wide range of low cross-correlation sequences of short length. This minimizes the sequence collisions within a given time-frequency resource block, while improving the detection probability at the receiver \cite{shahab2020grant,yuan2016multi,zte1608954,eid2017performance}. The spreading codes are also designed to simplify SIC at the receiver \cite{shahab2020grant}. Thus, MUSA is a suitable GF-NOMA multiple-access protocol for IoT-based \textcolor{black}{EMNs}.

\underline{\textit{Short Packet Communication for IoT-based \textcolor{black}{EMNs}:}}
To enable low-latency IoT communications, we use short packet communication by employing short frame structure and short packets \cite{sun2019optimizing}. Due to the short packet size and the low-latency requirements, the decoding errors in such communications cannot be ignored and the decoding probability cannot be characterized by Shannon's capacity without underestimating the latency and reliability \cite{sun2019optimizing}. In the finite block regime, the short packet transmission error probability has been characterized in the literature \cite{chen2021urllc}. 

\underline{\textit{State-of-the-Art in the Literature:}}
Next, we present the state-of-the-art in the literature related to MUSA-based GF-NOMA and short packet communications. Recently, several studies have proposed to use NOMA in short packet communications to reduce the latency of time-critical applications while improving the spectral efficiency \cite{sun2018short,han2019energy,hu2022optimization}. Moreover, the aforementioned merits of utilizing UAVs in the IoT infrastructure have motivated the modeling and analysis of UAV-assisted IoT networks with short packet communication to improve reliability, latency, throughput, and energy efficiency \cite{yu2020joint,pei2021joint,pan2019joint,chen2020power,wang2021packet,lei2022performance}.
Despite that the performance analysis of NOMA in IoT networks in the infinite block regime using stochastic geometry has also been widely investigated in literature \cite{zhang2020semi,wang2022uav,sreya2022adaptive}. There lacks analytical frameworks to evaluate the performance of MUSA-based short packet communication in UAV-assisted IoT networks. Prior studies on short packet communication have been mostly limited to the case of two NOMA users or a single transmitter-receiver pair \cite{zhang2022semi}, while the studies focusing on multiuser NOMA have not addressed short packet communication \cite{liu2021modeling}, latency \cite{fu2020block}, or UAV data aggregation \cite{abbas2017on}. For example, \cite{zhang2022semi} analyzed the outage performance in a semi-GF system where GF users and GB users share the same spectral resources in a two-user NOMA scenario. An analytical model of a GF-NOMA machine-type communication system was proposed in \cite{liu2021modeling} where the aggregate data rate without considering short packet decoding errors was analyzed. In \cite{fu2020block}, SIC decoding was designed based on power-domain NOMA which relies solely on the channel diversity of devices, while the latency requirements of short packets were not considered. The average system throughput under massive NOMA in the infinite and finite block length regimes was derived in \cite{abbas2017on}, where UAV-assisted data aggregation was not addressed. {To the best of our knowledge, there is no solution available to analyze and satisfy the high-reliability and low-latency requirements when an EoI occurs in a UAV-assisted IoT-based \textcolor{black}{EMNs}, which motivates our work.}

In this work, we consider a UAV-assisted IoT-based environmental monitoring system generating heterogeneous traffic with specific reliability and latency requirements. For this system, we develop an analytical framework to analyze the performance of short packet communication, and optimize the performance to satisfy reliability and latency requirements. The main contributions of this work are:
\begin{itemize}
\vspace{-1mm}
  \item We propose an adaptive transmission scheme for a UAV-assisted IoT-based \textcolor{black}{EMNs}, where IoT devices generate short packets with stringent latency and reliability requirements and produce heterogeneous traffic when an EoI occurs. The proposed scheme employs MUSA-based GF-NOMA, with ideal multi-user decoding (MUD) at the UAV, to enable IoT-to-UAV communications.
  \item We analyze the statistical performance of the MUSA-based GF-NOMA scheme and MUD at the UAV using the signal-to-interference-plus-noise ratio (SINR) coverage probability, which is defined as the average probability that a packet transmitted by a device during a given time slot in a transmission frame is not subjected to MUSA collision and the SINR exceeds a certain threshold. We also address the decoding error at the UAV due to transmission of short packets by considering short packet transmission error probability subject to packet size and transmission duration.
  \item We design an adaptive transmission scheme for the IoT-to-UAV communication by formulating and solving an optimization problem. We maximize the SINR coverage probability for a device with multiple packets to transmit, given a stringent latency constraint, by optimizing the number of transmission time slots, while satisfying the short packet reliability requirements with respect to short packet transmission error probability. The proposed scheme prevents the over-utilization of time-frequency resources under low-rate traffic and under-utilization of resources under high-rate traffic in an EoI.
  \item We compare our proposed scheme with a transmit power diversity scheme (TPDS) and a non-adaptive scheme (NAS). Our results show that the proposed scheme outperforms NAS and exhibits superior reliability and stability than the state-of-the-art TPDS at moderate to high average traffic rates. The results demonstrate the potential of UAV-assisted short packet communication with MUSA-based GF-NOMA to process high-rate traffic data when an EoI happens in an IoT-based \textcolor{black}{EMNs}.
\end{itemize}
\textit{Notations:} $\mathbb{P}(\cdot)$, $\mathbb{E}[\cdot]$, $\mathbb{Q}[\cdot]$, $\mathrm{G}[\cdot]$, $\mathrm{H}[\cdot]$, $\lfloor\cdot\rfloor$, and $\daleth$ denote the probability, expectation operator, $Q$-function, Barnes-G function, Harmonic function, Floor function, and Euler's constant, respectively. $f_{X}(x)$ and $F_{X}(x)$ denote the probability density function (PDF) and cumulative distributive function (CDF) of a random variable $X$, respectively. 

\section{System Model}\label{sec:system_model}
\begin{figure}[t]
  \centering
  \includegraphics[width=0.65\textwidth]{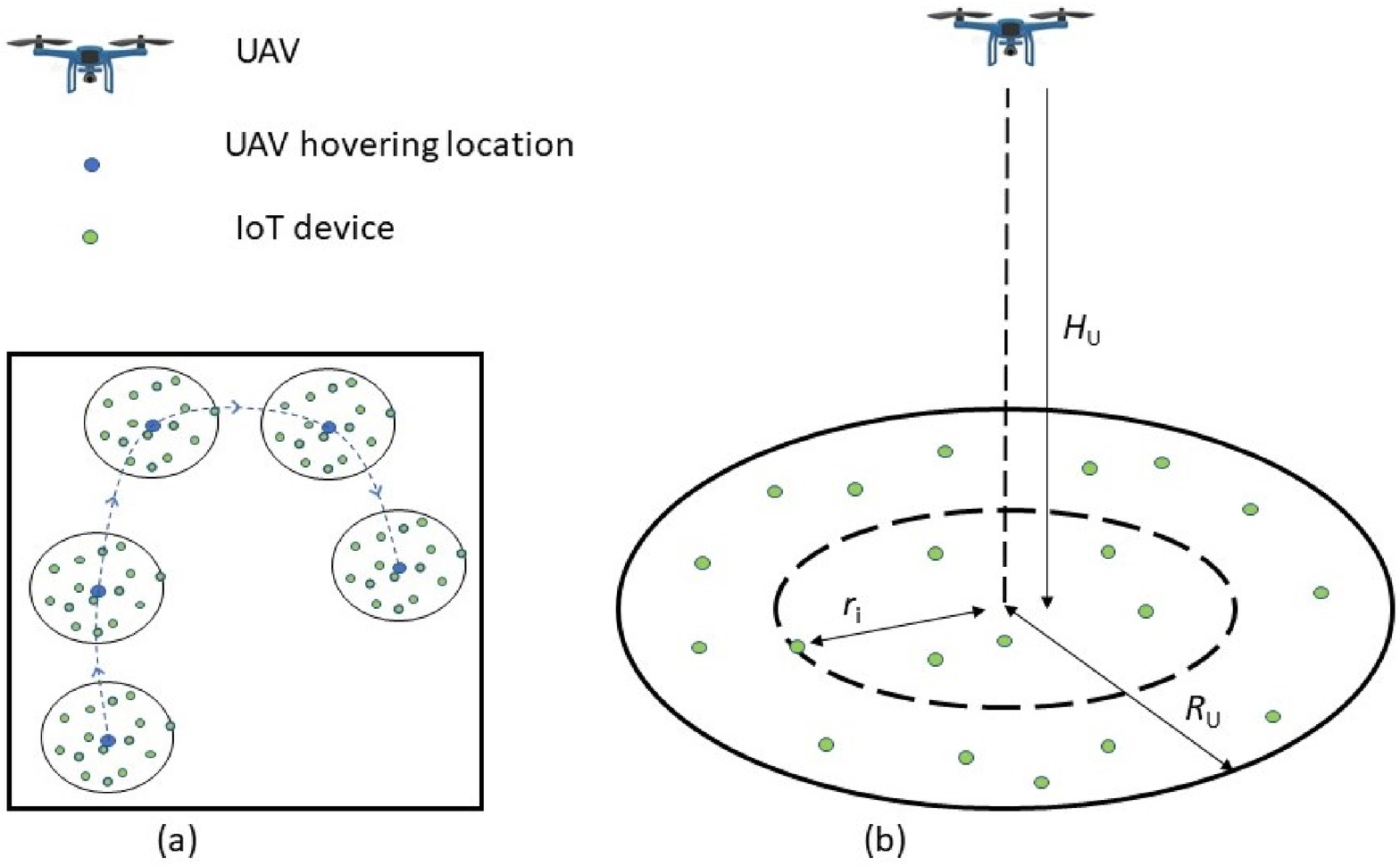}
  \caption{\textcolor{black}{Illustration of the system model. Fig. (a) shows different hovering points of the UAV's trajectory within the cell. The remote BS that is out of range of IoT devices is not shown in the figure as UAV-BS transmission is assumed to be highly reliable and fast and is not the main focus of this work.} Fig. (b) shows the serving zone area of the UAV at a given hovering point.}\label{fig:system_model}
\end{figure}

We consider an IoT-based environmental monitoring system where low-powered and short-range IoT devices are required to transmit under specific reliability and latency requirements to a remote terrestrial BS that is out of the range of the IoT devices, as depicted in Fig. \ref{fig:system_model}. To assist in the IoT-BS communication, a UAV is deployed as a data aggregator. The IoT devices that are located within the serving radius of the UAV transmit to the UAV and the successfully received data is aggregated and transmitted from the UAV to the BS. We assume that IoT-UAV and UAV-BS transmission phases are performed via orthogonal frequencies. Thus, there is no interference between these transmission phases. Also, the UAV uses a higher transmit power compared to the low-powered IoT devices. Furthermore, due to the high altitude of the UAV, it can establish high-quality communication with the BS. Thus, we assume highly reliable and fast communication between the UAV and the BS, and mainly focus on the IoT-UAV communication. 

We assume that the IoT devices generate low-rate smooth traffic during the majority of the time, and UAV is only capable of handling low-rate traffic under normal system settings, while satisfying predetermined reliability and latency constraints. However, when an EoI occurs, high-rate bursty traffic is generated by the IoT devices. Thus, the UAV needs to change its operating mode to accommodate the change in traffic conditions in order to satisfy the reliability and latency requirements. In the next subsections, we present the IoT system and UAV deployment models, traffic generation pattern of EoI, and the communication channel models.

\subsection{\underline{IoT System and UAV Deployment Model}}
The IoT devices are distributed on the ground, according to a homogeneous two-dimensional (2D) Poisson point process (PPP) with intensity $\omega$. We assume that the UAV is deployed in a fly-and-hover manner, following a predetermined trajectory with multiple hovering locations within the cell to aggregate data from the transmitting IoT devices and transmit the aggregated data to the BS\footnote{We assume that the UAV is initially located at the charging station and has sufficient energy to complete the communication tasks and return back to the charging station \cite{bushnaq2019aeronautical}.}. We also assume that the IoT-UAV data transmission only occurs at the hovering point\footnote{The UAV trajectory design and optimization has been addressed in various existing studies, e.g., \cite{zeng2017energy,wu2018joint,zeng2019energy}, and is outside the scope of this work.}. The transmission protocol is similar at every hovering location. Therefore, for the rest of this work, we only focus on the system performance at a single hovering location.

When the UAV arrives at the hovering location, it hovers at a fixed altitude $\hUAV$ for a fixed time duration $\hovtime$. At the given altitude, the serving zone of the UAV is defined as a disk of radius $\cellrad$. The hovering time duration $\hovtime$ consists of multiple transmission frames, each having a duration of $\frametime$. In each frame, the IoT devices located within the serving zone become active with an activation probability $\PAct$ and remain active for the entire frame duration $\frametime$. The transmitting devices are known as \textit{active users}. The distribution of the active devices in a frame is modeled as a thinned PPP distribution with radius $\cellrad$ with intensity $\denori=\omega\PAct$ \cite{haenggi2012stochastic}. Therefore, the number of active users in a transmission frame can be calculated as $\NAusers=\pi \cellrad^2 \omega \PAct$.


\subsection{\underline{Event-of-Interest Traffic Generation}}
We assume that the packets generated during the current frame are transmitted in the next frame. We also assume that an EoI can occur at any time. If an EoI occurs in the current frame, the system conditions to accommodate this change will occur in the next frame. We define the normal operating conditions and the EoI operating conditions of the IoT devices as the \textit{non-emergency} scenario and \textit{emergency} scenario, respectively, as described as follows:
\begin{definition}
  In the non-emergency scenario, each active device generates a single packet to transmit in the next frame.
\end{definition}
\begin{definition}
  In the emergency scenario where an EoI occurs, more devices become active and each active device generates multiple packets to transmit in the next frame. Specifically, each active device generates $\rho_{m}\sim\mathrm{Pois}(\nAVGpack)$ packets in the frame duration $\frametime$,
where $m\in\{1,2,\cdots,\NAusers\}$ and $\nAVGpack$ is the average number of packets generated by an active device.
\end{definition}

The IoT devices transmit short packets of size $\databits$ bits, and all packets must be transmitted in the next frame with the duration $\frametime$ which corresponds to the maximum allowable latency for transmission of packets\footnote{In this work, the transmission durations of the short packets transmitted by the IoT devices, e.g., those considered in Section \ref{sec:results}, conform to URLLC communication standards and the corresponding blocklengths are associated with the finite blocklength regime.}. We assume that the active device transmissions between two consecutive frames are independent of each other. Therefore, in the following we only focus on the transmission behavior in a single frame.

\subsection{\underline{Channel Model}}
We model the IoT-UAV channel link as a combination of large-scale path loss, which depends on distance and height, and small-scale Rayleigh fading\footnote{The studies \cite{perez2018on,arshad2018integrating,banagar2019gpp} used Rayleigh fading to model the small-scale fading for ground-to-air (G2A) communication. Our analytical framework is valid for both Nakagami-$m$ fading and Rayleigh fading models. However, due to the complexity of the analytical expressions in Section \ref{sec:performance}, the convexity of optimization problem presented in Section \ref{sec:OP} cannot be analytically expressed for Nakagami-$m$ fading. It has been shown in \cite{perez2018on,azari2017coexistence} that the quantitative performance trends in UAV communications remain unchanged even when the fading model is changed. For these reasons, we consider Rayleigh fading in this work.}. Particularly, we model its path loss as $\pathlossIOT=\PLcoef (r^2+\hUAV^2)^{\frac{\PLindex}{2}}$, where $r$, $\PLcoef$, and $\PLindex$ correspond to the horizontal distance between the UAV and the IoT device, path loss attenuation coefficient, and path loss exponent, respectively~\cite{zhou2019underlay}. The Rayleigh fading gain $\hfad$ follows an exponential distribution with unit mean. The PDF $f_{\hfad}(x)$ and the CDF $F_{\hfad}(x)$ of $\hfad$ are given by  $f_{\hfad}(x)=e^{-x}$ and $F_{\hfad}(x)=1-e^{-x}$, respectively. Moreover, the active IoT devices transmit to the UAV with fixed transmit power $\avgTransP=\frac{\Pmax}{\rho_{\mathrm{max}}}$, where $\Pmax$ is the transmit power budget of an IoT device and $\rho_{\mathrm{max}}=\max_{m}\{\rho_{m}\}, \forall m$ is the maximum number of packets generated by an IoT device in a given frame\footnote{We assume that the UAV broadcasts the transmit power information to the IoT devices at the beginning of the time frame.}.

\section{{Proposed UAV-assisted IoT Device Transmission Model}} \label{sec:IoT_UAV_trans}
In this section, we propose to use the MUSA-based GF-NOMA scheme for supporting the IoT-UAV transmission. The details are presented as follows.

\subsection{MUSA Spreading Code Generation}
MUSA is a code-domain multiple access scheme. In this work, we use a binary complex spreading code consisting of $(+1,-1)$ with length $J$. Therefore, each element of the complex spreading code is produced from the set $\{1+i,-1+i,-1-i,1-i\}$ before normalization. At the beginning of the transmission frame, the modulated symbols of each active device are spread by randomly chosen MUSA spreading codes, and the active devices’ spread symbols are superimposed and transmitted on the same time-frequency resource.

\subsection{IoT-UAV Transmission Frame Structure}
We consider that each transmission frame consists of $\nslot$ time slots with a duration $\slottime=\frac{\frametime}{\nslot}$. At the beginning of a frame, $\NAusers$ devices randomly select time slots from $\nslot$ available time slots. Depending on the scenario, the active devices either have one packet per device or multiple packets per device to transmit within the frame duration. We assume that an active device can only transmit a single packet in one time slot. Therefore, the time slots allocated for the packets of a given device are orthogonal. Once the time slots are selected, the devices randomly allocate MUSA sequences for the packets. Unlike slot selection, the MUSA sequence selected by a device is not orthogonal. Then, the active devices transmit their packets in randomly selected time slots using the corresponding randomly selected MUSA sequences. The transmissions between two consecutive time slots are independent of each other. Therefore, in the following, we present the transmission scheme and the recovery procedure for a given time slot.

\subsection{IoT-UAV Transmission Model}
For a given time slot with $\NAslot\leq\NAusers$ active devices, $\databits$ bits of each packet are modulated using binary phase shift keying to generate a stream of symbols for each device. After this, the modulated symbols of each device $i$, where $i\in\{1,2,\cdots,\NAslot\}$, is spread with a randomly chosen complex spreading code $s_{i}$ of length $J$ and transmitted over $J$ subcarriers. We denote $\mathbf{y}={[y_1,\cdots,y_{J}]}^{T}$ as the superimposed signal of undecoded devices where $[\cdot]^{T}$ is the transpose. Then, $\mathbf{y}$ is given by
\begin{equation}\label{eq:rec_sig3}
\mathbf{y}=\mathbf{G}\mathbf{P}^{\frac{1}{2}}\mathbf{x}+\mathbf{n},
\end{equation}
where $\mathbf{G}=[\mathbf{g}_1,\cdots,\mathbf{g}_{\NAslot}]$ is the equivalent channel of all the active devices, $\mathbf{x}=[x_1,\cdots,x_{\NAslot}]^{T}$ is the transmitted symbol vector, $\mathbf{P}^{\frac{1}{2}}=\textrm{diag}\left(\sqrt{\avgTransP},\cdots,\sqrt{\avgTransP}\right)\in \mathbb{R}^{\NAslot\times \NAslot}$ is the transmit power matrix and $\mathbf{n}\sim \mathcal{CN}(0,\sigma^2 \mathbf{I}_{J})$ is the noise vector. We then express $\mathbf{G}$ as $\mathbf{G}=\mathbf{H}\odot\mathbf{S}$, where $\mathbf{H}=[\mathbf{h}_1,\cdots,\mathbf{h}_{\NAslot}]$ with $\mathbf{h}_{i}=[h_{i,1},\cdots,h_{i,J}]^{T}$ being the channel gain vector for the $i$th device, $\mathbf{S}=[\mathbf{s}_1,\cdots,\mathbf{s}_{\NAslot}]$ with $\mathbf{s}_{i}=[s_{i,1},\cdots,s_{i,J}]^{T}$ being the MUSA sequence vector for the $i$th device, and $\odot$ is the element-wise product. We further denote $s_{i}\in S$ as the MUSA spreading code of length $J$ chosen by the $i$th device, and $S=\left\{s_{1},s_{2},\cdots,s_{N_{\nu}}\right\}$ as the pool of spreading codes with $N_{\nu}=9^{J}$ for binary MUSA spreading codes.

\subsection{Data Recovery at UAV}
At the UAV receiver, similar to prior studies \cite{wang2015comparison,eid2017performance,ameur2020power,oyerinde2019comparative}, the ideal minimum mean squared error (MMSE) successive interference cancellation (SIC) is implemented to recover the transmitted data packets. We use this ideal MMSE SIC receiver due to its ability to achieve the best possible sum rate for detecting multiple data streams \cite{zte1608954} and due to its low decoding complexity compared to other techniques such as message passing and maximum \textit{a posteriori} estimation \cite{ameur2020power}. This receiver operates under ideal assumptions that the UAV has prior knowledge on the number of active users, their MUSA signatures, and fading channels. As our analytical framework does not rely on the instantaneous knowledge of the active users nor instantaneous channel conditions, we can use ideal MMSE SIC to retrieve the transmitted data in a time slot\footnote{We note that, in reality, due to the GF nature of the access scheme, the UAV receiver may not have complete knowledge of the number active devices in the slot, their fading channels, and MUSA sequences. To address this practical concern, blind multi user detection (MUD) in MUSA is proposed \cite{zte164270,yuan2017blindAX}. The blind MUD uses blind estimation to select the device with the highest SINR by using blind channel estimation to reconstruct interference, and recover the transmitted symbols. Even though, the performance of blind MUD can approach that of ideal MMSE SIC, the implementation of which is relatively complex. Thus, we use ideal MMSE SIC due to its simplicity.}.

The SIC decoder uses the SINRs of the transmitting devices to estimate the transmitted symbols, reconstruct the interference, and decode the packets. This process is repeated until all the decodable packets are decoded. Moreover, we do not employ any power control mechanism at the IoT devices. Instead, the inherent received signal disparity due to the near-far effect and small-scale fading is utilized. Thus, the receiver utilizes both power disparity of the devices and low cross-correlation of the sequences to improve the decoding probability at the receiver. In the SIC decoding, the MMSE weights of the undecoded devices are computed at each iteration as \cite{ameur2020power}
\begin{equation}\label{eq:w_mmse} \mathbf{W}^{H}=\left(\mathbf{P}^{\frac{1}{2}}\mathbf{G}^{H}
\mathbf{G}\mathbf{P}^{\frac{1}{2}}+\sigma^2 \mathbf{\mathrm{I}}
  \right)^{-1} \mathbf{P}^{\frac{1}{2}}\mathbf{G}^{H},
\end{equation}
\noindent where $(\cdot)^H$ and $\mathbf{\mathrm{I}}$ correspond to the Hermitian transpose and identity matrix, respectively.

The undecoded devices are then sorted based on the distance, and the undecoded device with the minimum distance at the $k$th iteration, where $k\in\{1,\cdots,\NAslot\}$, is considered as the $k$th strongest device. The SINR of the $k$th strongest device is given by \cite{eid2017performance}
\begin{equation}\label{eq:pp_sinr}
\gamma_{k}=\frac{\avgTransP |w_{k}^{H}g_{k}|^2}
{\sum_{i=1+k}^{\NAslot} \avgTransP |w_{i}^{H}g_{i}|^2+\sigma^2 \|w_{k}^{H}\|},
\end{equation}
where $w_{k}$ is the $k$th MMSE weight. We consider that the $k$th strongest device is decoded if it is not subject to MUSA collision and $\gamma_{k}$ exceeds a given threshold $\threshold$. Otherwise, we consider that the $k$th strongest device and the rest weaker devices cannot be decoded. The decoded signal is reconstructed and its contribution to $\mathbf{y}$ is removed. This process is repeated until all the devices are decoded, provided there are no MUSA collisions and their SINRs exceed $\threshold$.

\subsection{Event-of-Interest Detection}

The existing transmission schemes and access mechanisms are designed based on overall average traffic rates, such that they over-utilize network resources under low-rate traffic but suffer from packet collisions under high-rate traffic. Thus, it is important to identify the occurrence of EoI so that the UAV-assisted IoT device transmission scheme is adapted to accommodate the change in traffic patterns while ensuring the efficient utilization of time-frequency resources. Therefore, in this subsection we propose the following approach to determine the occurrence of an EoI.

To determine if an EoI has occurred, the UAV performs the following procedure at the beginning of the current time frame. Prior to the transmission, the UAV knows the number of actives devices $\NAusers$ and the total number of packets $\rho_{\Sigma}$ to be transmitted in the frame\footnote{Note that active user detection has been widely addressed in the literature, which can be used by the UAV to know $\NAusers$.}. The UAV determines if an EoI has occurred by computing the average of number packets $\bar{\nAVGpack}=\frac{\rho_{\Sigma}}{\NAusers}$. If $\bar{\nAVGpack}>1$, the UAV determines the occurrence of an EoI and changes its operating mode in the emergency scenario. The main parameter that determines the severity of the emergency scenario is $\nAVGpack$ which is the rate of a Poisson distribution. $\nAVGpack$ can simply be approximated as $\bar{\nAVGpack}$. Alternatively, $\nAVGpack$ can be estimated by using the Multiple Hypothesis test. In this test,  $M-1$ independent hypotheses are tested as $\mathcal{H}_{2},\cdots,\mathcal{H}_{M}$, where $\mathcal{H}_{i}$ is the event that $\nAVGpack=i$, and the corresponding probability densities for $x$ conditioned on the given hypothesis is given as $p_{x|\mathcal{H}}(y|\mathcal{H}_{i})=\frac{i^{x}e^{-i}}{x!}$, where $x=\bar{\nAVGpack}$ is the calculated mean \cite{trees2001detection}. Then, the Maximum Likelihood Ratio test is conducted for $\mathcal{H}_{i}$ versus $\mathcal{H}_{i+1}$ to obtain $\hat{\nAVGpack}$ as
\begin{equation}\label{eq:likelihood_test}
  \Delta(x)=\frac{p_{x|\mathcal{H}}(y|\mathcal{H}_{i})}
  {p_{x|\mathcal{H}}(y|\mathcal{H}_{i+1})}
  \underset{\mathcal{H}_{i}}{\overset{\mathcal{H}_{i+1}} \gtrless} 1.
\end{equation}
The estimated $\nAVGpack$ is given by $\tau$, where $\mathcal{H}_{\tau}$ is the decision of the Multiple Hypothesis test.

The analytical framework presented in this work is applicable for any value of $\nAVGpack$ regardless of the estimation method. Thus, in the next section, we present an analytical framework to assess the performance of the system for the emergency scenario for a given $\nAVGpack$.

\section{Analytical Framework}\label{sec:performance}
The performance of the proposed UAV-IoT transmission protocol in Section \ref{sec:IoT_UAV_trans} depends on (i) the SINRs in \eqref{eq:pp_sinr}, (ii) the MUSA sequence collisions of the active devices within a transmission frame, (iii) the distance-based ordering of the undecoded devices, and (iv) decoding errors at the UAV due to short packet communication. While it may be possible to define a single metric that takes all these factors into account, we do not take this approach. Instead, for the sake of analytical tractability, we choose to define two metrics. The first metric accounts for the first three factors, while the second metric accounts for the short packet error probability. Such metrics allow us to leverage stochastic geometry to calculate the first metric and use the existing definition in the short packet communication literature to assess the second metric. Both metrics are then used to formulate the optimization problem in Section \ref{sec:OP}, for which we will develop an analytical solution. The definitions and results for the two metrics are presented in the next subsections.

\subsection{{SINR Coverage Probability}}
\begin{definition}
  The SINR coverage probability within a transmission frame, $\Psucc$, is defined as the average probability that a packet transmitted by a device in a given time slot within the transmission frame is not subjected to MUSA collision and the SINR exceeds a certain threshold $\threshold$.
\end{definition}

In a given time slot, the active devices with unique MUSA sequences are defined as \textit{singleton} devices, whereas the active users with colliding MUSA sequences are defined as \textit{collided} devices. We assume that the singleton devices are in SINR coverage if their SINRs exceed $\threshold$. On the other hand, we assume that the \textcolor{black}{collided devices are not in SINR coverage}. Thus, the SINR for a given singleton device, denoted as $\SINR$, is expressed as
\vspace{-2mm}
\begin{equation}\label{eq:SINR}
  \SINR=\frac{\avgTransP\PLcoef\left({\hat{r}^2+\hUAV^2}\right)^
  {-\frac{\PLindex}{2}}\hfad}{\Intsing+\Intsing+{\noiseP}},
\end{equation}
\noindent where $\hat{r}$ corresponds to the horizontal distance between a singleton device and UAV, and $\noiseP$ is the noise power, $\Intsing=\sum_{i\in\PPPsing}\avgTransP\PLcoef (r_{i}^{2}+\hUAV^2)^{-\frac{\PLindex}{2}}\hfad_{i}$ and $\Intcol=\sum_{j\in\PPPcol}\avgTransP\PLcoef (r_{j}^{2}+\hUAV^2)^{-\frac{\PLindex}{2}}\hfad_{j}$ denote the aggregate interference caused by the weaker singleton devices and all the collided devices, respectively, $\PPPsing$ represents the homogeneous PPP of the interfering singleton devices with intensity $\densing$, and $\PPPcol$ represents the homogeneous PPP of the collided devices with intensity $\dencol$. Both $\PPPsing$ and $\PPPcol$ are independent thinned PPP distributions of $\PPPori$ and the corresponding intensities can be derived using Lemmas $\ref{lemma:act_P}$ and $\ref{lemma:cf_P}$.

Next, we present useful Lemmas and Proposition to derive the SINR coverage probability $\Psucc$ defined in \eqref{eq:SINR}. First, we derive the probability that a packet of an active device chooses any time slot in Lemma \ref{lemma:act_P} and the probability that a random device transmitting in a given time slot is not subjected to MUSA sequence collision in Lemma \ref{lemma:cf_P}. Then, we characterize the singleton and collided interference in Lemmas \ref{lemma:Lap_Is} and \ref{lemma:Lap_Ic}, respectively. Finally, we present the expression for the SINR coverage probability $\Psucc$ in Theorem \ref{theo:dec_prob}.

\begin{lemma}\label{lemma:act_P}
  Given that the average number of packets generated by an active device is $\nAVGpack$, and the number of time slots available for transmission is $\nslot$, the probability that a packet of an active device chooses any time slot, denoted by $P_{\nAVGpack}$, is given by
\vspace{-4mm}
\begin{equation}\label{eq:P_act}
\Plambda=\sum_{L=1}^{\nslot}\frac{e^{-\nAVGpack}
{\nAVGpack}^{L}}{\nslot (L-1)!}
+\sum_{L=\nslot+1}^{\nAVGpack+L_{\mathrm{lim}}}\frac{e^{-\nAVGpack}
{\nAVGpack}^{L}}{L!}.
\end{equation}

\noindent \begin{IEEEproof}
$\Plambda$ is expressed as
\vspace{-2mm}
  \begin{equation}\label{eq:act_prob_proof}
    \Plambda=\sum_{L}P_{\nAVGpack|L} f_{\nAVGpack}(L),
      \vspace{-3mm}
  \end{equation}
\noindent where $P_{\nAVGpack |L}$ is the conditional probability of a device choosing any time slot from $\nslot$ available time slots to transmit one of its packets, given that the device has $L\sim\textrm{Pois}(\nAVGpack)$ number of packets to transmit within the time frame. Specifically, $P_{\nAVGpack|L}$ is given by
\vspace{-4mm}
  \begin{equation}\label{eq:act_prob_cond}
    P_{\nAVGpack|L}=
    \begin{cases}
      \frac{L}{\nslot}, & \mbox{if } L<\nslot \\
      1, & \mbox{otherwise},
    \end{cases}
      \vspace{-3mm}
  \end{equation}

\noindent where $f_{\nAVGpack}(L)=e^{-\nAVGpack}\frac{{\lambda}^{L}}{L!}$ is the PDF of the number of packets generated by an active device.
\end{IEEEproof}
\end{lemma}

\begin{lemma}\label{lemma:cf_P} 
  The collision free probability, $\Pcf$, which is the probability that an active device in a given time slot does not choose the same MUSA sequence as another active device in the same time slot, is given by
  \begin{align}\nonumber
    \Pcf&=(1-\Plambda)^{\NAusers-1}+
    \sum_{n=1}^{\NAusers-1}\binom{\NAusers-1}{n}{\Plambda}^{n}\\ \label{eq:Pcf}
    &\times{(1-\Plambda)}^{\NAusers-n-1}
    {\left(\frac{\Nmusa-1}{\Nmusa}\right)}^{n}.
  \end{align}
  \normalsize

\noindent  \begin{IEEEproof}
For a device to be collision free in a given time slot, either no other device transmits on the same time slot or the devices that transmit on the same time slot must not choose the same MUSA sequence as the considered device. Thus, the collision free probability of a device can be expressed as $\Pcf=P_{o}+P_{\sum}$ where, $P_{o}=(1-\Plambda)^{\NAusers-1}$ is the probability that only one device, out of $\NAusers$ devices, is active and all the other $\NAusers-1$  devices are not active in the given time slot. Moreover, $P_{\sum}$ is the probability that given a device and $n\in \{1,\cdots,\NAusers-1\}$ other devices are active in the given time slot, there is no MUSA sequence collision between the considered device and other $n$ active devices. We then derive $P_{\sum}$ as follows:


The probability that $n$ devices transmit in a given time slot is given by $(\Plambda)^{n}(1-\Plambda)^{\NAusers-1-n}\binom{\NAusers-1}{n}$ and the probability that $n$ devices do not choose the same MUSA sequence as the considered device in the same time slot is given by ${\left(\frac{\Nmusa-1}{\Nmusa}\right)}^{n}$. Therefore, $P_{\sum}$ can be derived as
\small
\begin{equation}\label{eq:P_sum}
  P_{\sum}=\sum_{n=1}^{\NAusers-1}\binom{\NAusers-1}{n} (\Plambda)^{n} (1-\Plambda)^{\NAusers-n-1} {\left(\frac{\Nmusa-1}{\Nmusa}\right)}^{n}.
\end{equation}
\end{IEEEproof}
\end{lemma}
\normalsize

Using Lemmas \ref{lemma:act_P} and \ref{lemma:cf_P}, the number of singleton devices $\Nsing$, in a given time slot can be computed as $\Nsing=\NAusers \Plambda \Pcf$. Next, we characterize the distance distribution and the interference on the singleton devices. First, we present an assumption for the purpose of analytical tractability.

\begin{assumption} \label{assum:order}
The impact of path loss is more dominant compared to that of small-scale fading on the received signal power. Thus, for analytical tractability, we consider the fixed \textcolor{black}{NOMA decoding order} strategy based on the distance of devices from the UAV, as opposed to the dynamic \textcolor{black}{NOMA decoding order} strategy based on the instantaneous received signal power. Therefore, the distance-based ordering of $\Nsing$ singleton devices is $r_{1}\geq r_{2} \geq \cdots r_{k}\geq \cdots {r}_{\Nsing}$, where $r_{1}$ is the horizontal distance of the strongest device that \textcolor{black}{would be decoded} first and ${r}_{\Nsing}$ is that of the weakest device that \textcolor{black}{would be decoded} last. It follows that for the $k$th strongest device with $\hat{r}=r_{k}$, the interference from the singleton devices is only generated by devices located farther away from $\hat{r}$ from the UAV \cite{tabassum2017modeling}.
\end{assumption}

Next, we characterize the distance distribution of the $k$th strongest singleton device in Lemma \ref{lemma:dist} based on Assumption \ref{assum:order}.

\begin{lemma}\label{lemma:dist}
The distance distribution of the $k$th strongest singleton device located at a horizontal distance $\hat{r}$ away from the UAV is given by
\small
  \begin{equation}\label{eq:dist_pdf_kth}
    f_{\hat{r}}(x)=\frac{\Nsing!}{(j-1)!(\Nsing-j)!}
\left(\frac{2\hat{r}^{2j-1}}{\cellrad^{2j}}\right)
{\left(1-\frac{{\hat{r}}^2}{\cellrad^2}\right)}^{\Nsing-j},
  \end{equation}
  \normalsize
\noindent where $\Nsing$ is the number of singleton devices in the time slot.

\begin{IEEEproof}
Based on the $k$th order statistics, the PDF of the distance distribution $r_{k}=\hat{r}$ of the $k$th strongest device is expressed as
\vspace{-3mm}
  \begin{equation}\label{eq:pdf_dist_k}
    f_{\hat{r}}(x)=\frac{\Nsing!}{(k-1)!(\Nsing-k)!} f_{r}(x) (F_{r}(x))^{k-1}(1-F_{r}(x))^{\Nsing-k},
  \end{equation}
\noindent where $f_{r}(x)=\frac{2x}{\cellrad^2}$ and $F_{r}(x)=\frac{x^2}{\cellrad^2}$ are the PDF and CDF of the distance distribution of a device located at a horizontal distance of $x$ away from the UAV serving zone of radius $\cellrad$, respectively.
\end{IEEEproof}
\end{lemma}

For a given time slot, to characterize the aggregated interference $\Intsing+\Intcol$ at the $k$th strongest singleton device located at a horizontal distance of $\hat{r}$ from the UAV, we compute the Laplace transforms of $\Intsing$ and $\Intcol$ at $s$, conditioned on the random distance $\hat{r}$, which we denote by $\mathcal{L}_{\Intsing}(s)$ and $\mathcal{L}_{\Intcol}(s)$, respectively.

\begin{lemma}\label{lemma:Lap_Is} 
  The Laplace transform of interference on the $k$th strongest singleton device generated by the weaker singleton devices, denoted by $\mathcal{L}_{\Intsing}(s)$, is derived as
  \small
  \begin{equation}\label{eq:lap_sing}
    \mathcal{L}_{\Intsing}(s)=\exp\left(-2\pi \denori \Pcf\int_{\hat{r}}^{\cellrad}
    {\left(1-\frac{1}{1+s\avgTransP\PLcoef (r^2+{\hUAV}^2)^{-\frac{\PLindex}{2}}}
    \right)}r dr
    \right).
  \end{equation}
  \normalsize

\begin{IEEEproof}
Please see Appendix \ref{app:laplace_Is}.
\end{IEEEproof}
\end{lemma}

\begin{lemma} \label{lemma:Lap_Ic}
  The Laplace transform of interference on the $k$th strongest singleton device generated by the collided devices, denoted by $\mathcal{L}_{\Intcol}(s)$, is derived as
  \begin{align}\nonumber
    \mathcal{L}_{\Intcol}(s)&=\exp\Bigg(-2\pi \denori (1-\Pcf)\int_{0}^{\cellrad}
    \Bigg(1 \\ \label{eq:lap_col}
    &-\frac{1}{1+s\avgTransP\PLcoef (r^2+{\hUAV}^2)^{-\frac{\PLindex}{2}}}
    \Bigg)r dr
    \Bigg).
  \end{align}
  \normalsize

\begin{IEEEproof}
The proof of \eqref{eq:lap_col} is similar to that of \eqref{eq:lap_sing}, except that the collided devices form a thinned PPP distribution of intensity $\dencol=\denori (1-\Pcf)$ within a disk of radius $\cellrad$. Therefore, the lower and upper integral limits in Campbell's theorem correspond to $0$ and $\cellrad$, respectively.
\end{IEEEproof}
\end{lemma}

Next, we derive the probabilities associated with the \textcolor{black}{SINR coverage of the singleton devices. We present the conditional probability that the $k$th strongest singleton device is in SINR coverage given the stronger singleton devices are in SINR coverage in Proposition \ref{prop:cond_prob}.} Then, we present the SINR coverage probability within a transmission frame in Theorem \ref{theo:dec_prob}.

\begin{proposition}\label{prop:cond_prob}
In a given time slot, the conditional probability that the $k$th strongest singleton device located at a horizontal distance of $\hat{r}$ away from the UAV is in SINR coverage, given that $k-1$ stronger singleton devices are in SINR coverage, is derived as
  \begin{align}\nonumber
    P_{COND}^{k}&=\int_{0}^{\cellrad}\exp(-s\noiseP)\mathcal{L}_{\Intsing}(s)
    \mathcal{L}_{\Intcol}(s)
     \frac{\Nsing!}{(j-1)!(\Nsing-j)!}\\ \label{eq:prob_sic}
  &\times\left(\frac{2\hat{r}^{2j-1}}{\cellrad^{2j}}\right)
{\left(1-\frac{{\hat{r}}^2}{\cellrad^2}\right)}^{\Nsing-j} d\hat{r},
  \end{align}
\noindent where $s=\frac{\threshold {({\hat(r)}^2+\hUAV^2)}^{\frac{\PLindex}{2}}}{\avgTransP\PLcoef}$ and $\Nsing$ is total number of singleton devices in a given time slot.

\begin{IEEEproof}
$P_{COND}^{k}$ can be expressed as
\begin{equation}\label{eq:Psic_proof}
  P_{COND}^{k}=\int_{0}^{\cellrad} P_{SIC|\hat{r}}^{k}(s) f_{\hat{r}}(\hat{r})d\hat{r},
\end{equation}
\noindent where $P_{COND|\hat{r}}^{k}$ is the conditional probability that the $k$th singleton device is in SINR coverage given it is located at $\hat{r}$ distance from the UAV. To further derive \eqref{eq:Psic_proof}, we obtain $P_{SIC|\hat{r}}^{k}$ as
\small
\begin{subequations}\label{eq:Psic_cond_proof1}
  \begin{align}\nonumber
  P_{SIC|\hat{r}}^{k}(s)
  &=\mathrm{Pr}\left(\frac{\avgTransP\PLcoef \left({{\hat{r}}^{2}+{\hUAV}^2}\right)^{-\frac{\PLindex}{2}}\hfad}
  {\Intsing+\Intcol+\noiseP}\geq \threshold\right)\\ \nonumber
  &=\mathbb{E}_{\Intsing,\Intcol}\left[1-F_{\hfad}\left(\frac{\threshold \left({{\hat{r}}^{2}+{\hUAV}^2}\right)^{-\frac{\PLindex}{2}}}
  {\avgTransP\PLcoef}(\Intsing+\Intcol+\noiseP)\right)\right]\\
  \label{eq:Psic_proof_a}
  &=\mathbb{E}_{\Intsing,\Intcol}\left[\exp\left(-s(\Intsing+\Intcol
  +\noiseP)\right)\right]\\ \nonumber
  &=\exp(-s\noiseP)\mathbb{E}_{\Intsing}\left[\exp(-s \Intsing)\right]
  \mathbb{E}_{\Intcol}\left[\exp(-s \Intcol)\right]\\
  \label{eq:Psic_proof_b}
  &=\exp(-s \noiseP)\mathcal{L}_{\Intsing}(s)\mathcal{L}_{\Intcol}(s),
  \end{align}
\end{subequations}
\normalsize
\noindent where \eqref{eq:Psic_proof_a} is obtained using the CDF of the small-scale fading gain $\hfad$. By substituting \eqref{eq:lap_sing}, \eqref{eq:lap_col}, \eqref{eq:Psic_cond_proof1}, and \eqref{eq:dist_pdf_kth} to \eqref{eq:Psic_proof}, we obtain \eqref{eq:prob_sic}.
\end{IEEEproof}
\end{proposition}

  \begin{figure*}[t] 
    \begin{align}\nonumber
    \Psucc&=\sum_{k=1}^{\Nsing}\frac{\nslot}{\NAusers\nAVGpack}
    \prod_{j=1}^{k}\int_{0}^{\cellrad}e^{-s\noiseP}\exp\left(-2\pi \denori \Pcf\int_{\hat{r}}^{\cellrad}
    {\left(1-\frac{1}{1+s\avgTransP\PLcoef (r^2+{\hUAV}^2)^{-\frac{\PLindex}{2}}}
    \right)}r dr
    \right)\\ \label{eq:dec_prob}
    &\exp\left(-2\pi \denori (1-\Pcf)\int_{0}^{\cellrad}
    {\left(1-\frac{1}{1+s\avgTransP\PLcoef (r^2+{\hUAV}^2)^{-\frac{\PLindex}{2}}}
    \right)}r dr \right)
     \times\frac{\Nsing!}{(j-1)!(\Nsing-j)!}
   \left(\frac{2\hat{r}^{2j-1}}{\cellrad^{2j}}\right)
   {\left(1-\frac{{\hat{r}}^2}{\cellrad^2}\right)}^{\Nsing-j} d\hat{r},
  \end{align}
  \rule{18.2cm}{0.5pt}
  \end{figure*}
  \normalsize
\noindent where $s=\frac{\threshold \hat{r}^{\PLindex}}{\avgTransP\PLcoef}$.

Next, we present another assumption made for the sake of tractable analysis.

\begin{assumption}\label{assum:joint_prob}
In a given time slot, for a singleton device to be in SINR coverage, the singleton devices with the stronger average received powers, i.e., singleton devices that are closer to the UAV than the current specified singleton device, must be in SINR coverage. Since the singleton devices are independently distributed in $\cellrad$ and the small-scale fading is exponentially distributed, the SINR coverage events of singleton devices are assumed to be independent.
\end{assumption}

We now derive and present $\Psucc$ in the following theorem.

\begin{theorem}\label{theo:dec_prob}
  The SINR coverage probability within the transmission frame, $\Psucc$, which is the average probability that a packet transmitted by a device in a given time slot within the frame is not subjected to MUSA collision and the SINR exceeds a certain threshold $\threshold$, is given by  \eqref{eq:dec_prob} at the top of this page.

\begin{IEEEproof}
Based on Assumption \ref{assum:joint_prob}, the probability that the $k$th strongest device is in SINR coverage is given as $P_{\mathrm{COV}}^{k}=\prod_{j=1}^{k}P_{COND}^{j}$. The probability that the devices are in SINR coverage within the transmission frame is given by $\frac{\sum_{k=1}^{\Nsing}P_{\mathrm{COV}}^{k}\nslot}{\NAusers\nAVGpack}$. Thus, substituting \eqref{eq:Psic_cond_proof1} into this, we obtain \eqref{eq:dec_prob}.
\end{IEEEproof}
\end{theorem}

\begin{remark}\label{rem:Psucc}
For a given $\nslot$, when the number of devices $\NAusers$ and the average number of packets $\nAVGpack$ increase, the SINR coverage probability within the transmission frame decreases. On the other hand, for a given $\NAusers$ and $\nAVGpack$, when $\nslot$ increases, the SINR coverage probability within the transmission frame increases.
\end{remark}

\begin{remark}
We note that $\Psucc$ in \eqref{eq:dec_prob} does not depend on the maximum transmit power, as we assume that all the active devices in the transmission frame transmit with equal power $\avgTransP$. As such, $\Psucc$ gives the lower bound on the performance, due to lower receiver power diversity.
\end{remark}

\subsection{Short Packet Transmission Error Probability}

To address the decoding errors at the UAV due to short packet communications in the analysis, we define the short packet transmission error probability, denoted by $\errorProb$, as follows.
\begin{definition}
  In the finite block regime, the short packet transmission error probability, $\errorProb$, is defined as the error probability of receiving $\databits$ bits of data within a single time slot duration $\slottime=\frac{\frametime}{\nslot}$, when the SINR of the device is $\gamma$. Here, $\errorProb$ characterizes the reliability of the system, given by \cite{chen2021urllc}
\begin{equation}\label{eq:error_const_PA}
\errorProb=\mathbb{Q}\left[\sqrt{\frac{\bandwidth \frametime}{V \nslot}}
\left(\log_{2}{(1+\gamma)-\frac{\databits\nslot}{\bandwidth \frametime}}
\right)\right],
\end{equation}
\noindent where $\mathbb{Q}\left[\cdot\right]$ is the $Q$-function, $\bandwidth$, $\frametime$, $\databits$, $V$, and $\nslot$ represent the bandwidth, frame duration which is the maximum allowable latency, size of the short packet transmitted in a time slot, channel dispersion, and the number of time slots, respectively.
\end{definition}

\begin{remark}\label{rem:error}
Given a device which has an SINR $\gamma$ and packet size of $\databits$, its transmission reliability increases when the transmission duration increases. Thus, we see from \eqref{eq:error_const_PA} that $\errorProb$ decreases when the number of time slots $\nslot$ within the frame increases.
\end{remark}

It can be seen from Remarks \ref{rem:Psucc} and \ref{rem:error} that both SINR coverage probability $\Psucc$ and short packet decoding error probability $\errorProb$ depend on the number of time slots $\nslot$ within a transmission frame. Therefore, in an EoI both $\Psucc$ and $\errorProb$ must be considered simultaneously when UAV changes its operating mode to accommodate the change in traffic conditions, for satisfying the relevant reliability and latency constraints. Thus, in the next section, we propose an adaptive transmission scheme to prevent the over-utilization of resources under low-rate traffic and under-utilization of resources under high-rate traffic, by optimizing the number of time slots in a transmission frame.

\section{Proposed Adaptive Transmission Scheme}\label{sec:OP} %
In this section, we propose an adaptive transmission scheme for IoT-UAV communication by optimizing the number of time slots to maximize the SINR coverage probability of active devices under stringent reliability constraint (i.e., maximum short packet transmission error probability) and latency constraint in an EoI.

From \eqref{eq:dec_prob}, we can see that the SINR coverage probability is a function of the MUSA collision probability and the SINR. The MUSA collision probability is inversely proportional to the number of time slots in a frame. When the number of time slots increases, the average number of devices with colliding MUSA sequence increases. This decreases the number of singleton devices that can be recovered via SIC decoding and decreases the SINR coverage probability. Moreover, when the number of singleton devices in a time slot increases, the interference on the stronger singleton devices caused by the weaker singleton devices increases. This decreases the SINR and decreases the SINR coverage probability. On the other hand, when the number of time slots increases, the transmission time of a packet decreases and thus, the short packet transmission reliability decreases. Clearly, there is a trade-off between having less collisions and achieving high transmission reliability. To address this trade-off, we define the optimization problem as ``\emph{What is the optimal number of time slots that maximizes the SINR coverage probability of an active device having multiple packets to transmit within a given time duration $\frametime$, while satisfying the reliability constraints of short packet transmission?}'' \textcolor{black}{We formulate this optimization problem with design parameter $\nslot$, which is then used to reveal optimal $\slottime$, as follows:}
\begin{subequations}\label{eq:OP_main}
\begin{align}\nonumber
\ifCLASSOPTIONdraftclsnofoot \vspace{-2mm} \fi
\mathbf{OP}: &\underset{\nslot}{\text{min }} (-\Psucc) \\
\text{s.t.~}
\mathrm{C}_1:& \nslot-(\NAusers\nAVGpack+\delta(\nAVGpack)) \leq 0, \\
\mathrm{C}_2:& \nAVGpack-\nslot \leq 0,\\ \label{eq:c3}
\mathrm{C}_3:& \mathbb{Q}\left[\sqrt{\frac{\bandwidth \frametime}{V\nslot}}
  \left(\log_{2}{(1\!+\!\gamma)-\frac{\databits\nslot}{\bandwidth \frametime}}
  \right)\right]\!-\!\epsilon \leq 0\\
\mathrm{C}_4:& \lambda_{\mathrm{min}}-\nAVGpack \leq 0, \\
\mathrm{C}_5:& \nAVGpack-\lambda_{\mathrm{max}} \leq 0, \\
\mathrm{C}_6:& R_{\mathrm{min}}-\cellrad \leq 0,
\ifCLASSOPTIONdraftclsnofoot \vspace{-2mm} \fi
\end{align}
\end{subequations}

\noindent where $\nslot=\frac{\frametime}{\slottime}$, with $\frametime,\slottime\in\mathbb{R}$. Thus, $\nslot\in\mathbb{R}$. However, in reality $\nslot\in\mathbb{Z}^{+}$. Therefore, to solve this OP, we relax this constraint and consider $\nslot\in\mathbb{R}$, and then use $\tilde{\nslot}=\lfloor\nslot^{*}\rfloor$, where $\nslot^{*}$ is the solution to the $\textbf{OP}$, and $\tilde{\nslot}$ is the practical value of number of time slots. $C_{1}$ defines an approximate upper bound on the number of time slots. While this upper bound ensures the efficient use of resources, $\delta(\nAVGpack)$ in $C_{1}$ ensures that the resources are not overused, i.e., lower $\delta(\nAVGpack)$ allows a higher upper bound and vice versa. $C_{2}$ defines the lower bound where the number of time slots must be greater than the average number of packets generated by a device. This ensures the minimal packet drop at the UAV. $C_{3}$ mandates that the error in short packet transmission satisfies the reliability constraint, where $\epsilon$ in \eqref{eq:c3} is the maximum allowable error in short packet transmission and $V$ is approximated as $(\log_{2}(e))^2$ \cite{chen2021urllc}. \textcolor{black}{Moreover, $C_{3}$ ensures that the successful transmission of a packet is completed within a maximum allowable latency of $\frametime$.} In $C_{3}$, we assume that the minimum time slot duration is associated with the maximum SINR, as the time slot duration is inversely proportional to the SINR for a given $\epsilon$. We also assume that the maximum SINR corresponds to the scenario where the device closest to the UAV is the only active device in a given time slot. Thus, $\gamma$ in \eqref{eq:c3} is approximated by signal-to-noise ratio for mathematical tractability. $C_{4}$ defines the minimum $\nAVGpack$ for mode transition in an emergency scenario. $C_{5}$ defines the worst case practical value for $\nAVGpack$. $C_{6}$ is the minimum serving radius of the UAV aggregator.

The objective function and the constraints in \eqref{eq:OP_main} satisfy the second-order constraints for convexity \cite{boyd2004convex}. Thus, $\mathbf{OP}$ in $\eqref{eq:OP_main}$ is a convex optimization problem. Therefore, the optimal solution to the $\mathbf{OP}$ can be obtained by finding the solution that satisfy the relevant Karush-Kuhn-Tucker (KKT) conditions.


\begin{theorem}\label{theo:OP}
\textcolor{black}{
The practical number of time slots $\tilde{\nslot}$, which will maximize $\Psucc$ within the frame duration $\frametime$ in an emergency scenario where the active devices generate $\nAVGpack$ packets in average is given by
\begin{equation}\label{eq:opt_ns}
\tilde{n}_{\mathrm{s}}=\lfloor\nslot^{\ast}\rfloor=\lfloor\min\left\{{\nslot}_{\nAVGpack},
n_{\epsilon}\right\}\rfloor,
\end{equation}
\noindent where, $\nslot^{\ast}$ is the solution to the $\mathbf{OP}$ in \eqref{eq:OP_main}, ${\nslot}_{\nAVGpack}=\NAusers\nAVGpack+\delta(\nAVGpack)$ and $n_{\epsilon}$ is given by the solution to the following equation
\begin{equation}\label{eq:op_sol1}
\mathbb{Q}\left[\sqrt{\frac{\bandwidth \frametime}{n_\errorProb}}
\left(\ln{(1+\gamma)-\frac{\databits}{\bandwidth \frametime}}\ln(2) n_\errorProb\right)\right]-\errorProb=0.
\end{equation}
}

\begin{IEEEproof}
Please see Appendix \ref{app:OP_sol}.
\end{IEEEproof}
\end{theorem}

\begin{remark}
The average number of packets $\nAVGpack$ provides a lower bound on the optimal design variable on the optimization problem in \eqref{eq:OP_main}. We note that $\min{\left({\nslot}_{\nAVGpack}, n_{\errorProb}\right)}$ gives the \textcolor{black}{optimal design parameter $\nslot^{\ast}$} when the all the constraints are slack, except for the first constraint. Particularly, ${\nslot}_{\nAVGpack}$ is the optimal number of time slots when the traffic is low. On the other hand, $n_{\errorProb}$ is the optimal number of time slots when the traffic is relatively high. This is due to the fact that in the emergency scenario where $\nAVGpack$ is high, the upper bound on the allowable number of time slots is bounded by the reliability constraint associated with short packet transmissions.
\end{remark}

\begin{remark}
The optimal solution to $\mathbf{OP}$ depends on the statistical channel information, the average number of packets $\nAVGpack$ which can be estimated, and the active number of devices $\NAusers$ which is known to the UAV at the beginning of the frame. Hence, the UAV can solve $\mathbf{OP}$ for different $\NAusers$ and $\nAVGpack$ before the beginning of transmission. Note that the UAV is only required to solve the $\mathbf{OP}$ when an EoI occurs.
\end{remark}

\section{Numerical Results}\label{sec:results}
In this section, we first evaluate the performance of the system in an emergency scenario under the non-optimal setting. The performance evaluation is conducted with respect to the number of active users $\NAusers$, severity of the emergency scenario which correlates to the average number of packets $\nAVGpack$ generated by a user, and the number of available time slots $\nslot$ to transmit within the given latency constraint $\frametime$ seconds. Next, we compare the performance of the proposed scheme presented in Section \ref{sec:OP} to several benchmarks.
\small
\begin{table}[t]
\centering
  \caption{Parameter values for numerical and simulation results.}\label{table:parameter_val}
\begin{tabular}{|l|l|l|}
\hline
\textbf{Parameter}      & \textbf{Symbol} & \textbf{Value} \\ \hline
UAV serving zone radius &    $\cellrad$             & $50$ m   \\ \hline       SINR threshold           &     $\threshold$            & $0$ dB         \\ \hline
Time frame duration     &         $\frametime$        & \textcolor{black}{$10^{-3}$} s        \\ \hline
 Pathloss coefficient     & $\PLcoef$                & $0$ dB         \\ \hline
Packet size             &   $\databits$              & \textcolor{black}{$200$ bits}      \\ \hline
 Pathloss exponent        &   $\PLindex$              & $2.2$          \\ \hline
Bandwidth               &             $\bandwidth$    & \textcolor{black}{$5$ MHz}         \\ \hline
 Noise power              &      $\noiseP$           & $-100$ dBm     \\ \hline
Number of MUSA sequences &          $\Nmusa$       & $64$      \\ \hline
Maximum transmit power   &           $\Pmax$      & $10$ dBm       \\ \hline
\end{tabular}
\normalsize
\end{table}
\normalsize

The simulation results for the MUSA-based GF-NOMA with ideal MMSE MUD scheme are obtained using system level computer simulations in Matlab where we average over $10^4$ Monte Carlo simulation runs. The analytical results are obtained using \eqref{eq:dec_prob} and the proposed scheme is obtained using \eqref{eq:opt_ns}. The numerical integrations in \eqref{eq:dec_prob} are evaluated using Mathematica or approximated with Simpson's integral. In the $\mathbf{OP}$, the gamma function, harmonic number function, Barnes-G function, and $Q$-function are evaluated numerically using Mathematica. The parameter values used for the results are presented in Table \ref{table:parameter_val}.  We clarify that these values are consistent with other relevant studies in the literature, e.g., \cite{senadhira2020uplink,ren2020joint,she2019ultra}. For the purpose of generating the results, we consider a fixed UAV altitude $\hUAV=125$ m, and assume that the IoT-UAV communication link has unit probability of line-of-sight at all times \cite{gppTR36777}.

\begin{figure*}[t]
\centering
\subfigure[$\Psucc$ vs. $\nAVGpack$ for different $\NAusers=10,15,20$. ]{\label{fig:model_vali}\includegraphics[width=0.45\textwidth]{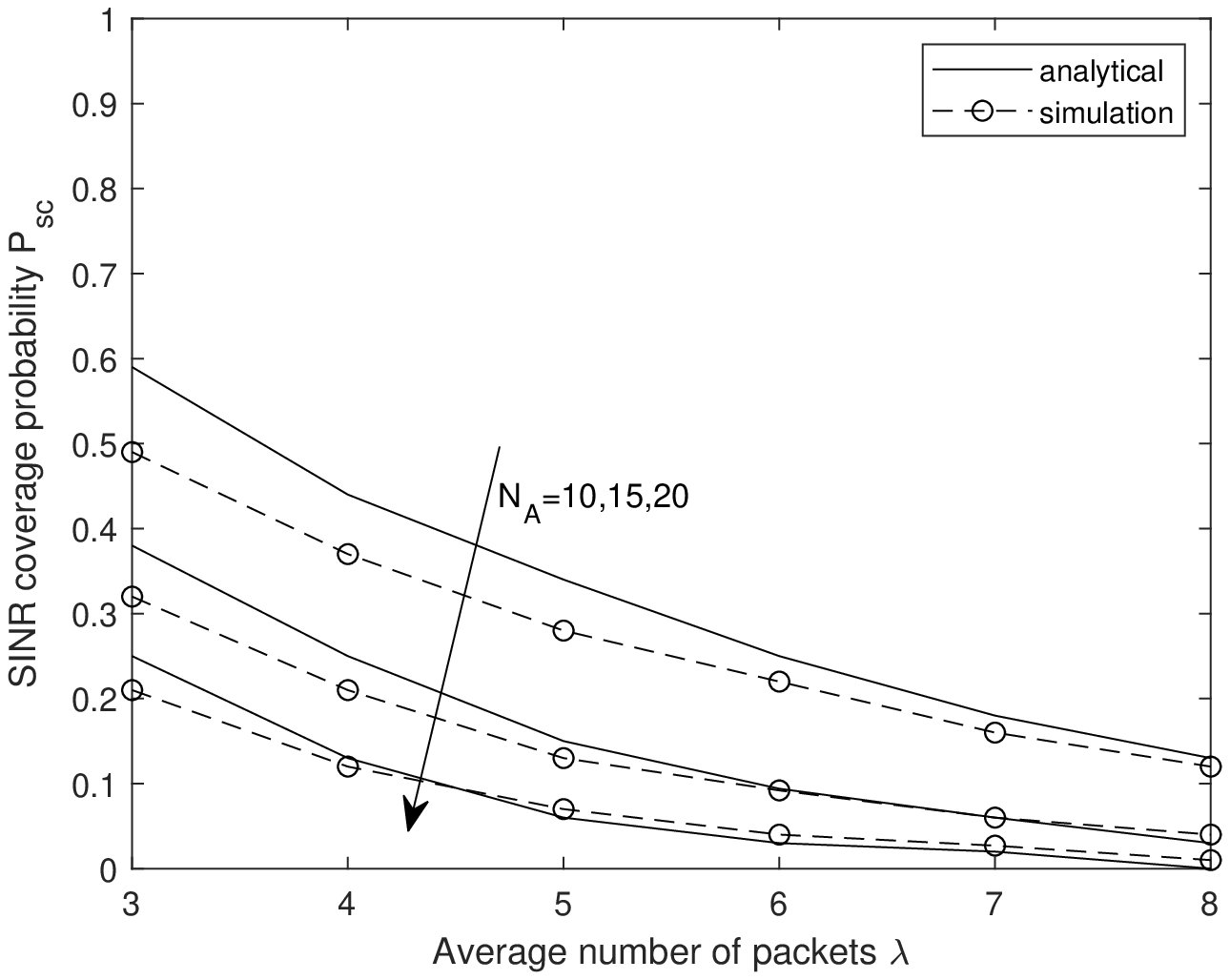}}
\subfigure[$\Psucc$ vs. $\NAusers$ for different $\nAVGpack=2,4,6,8,10$. ]{\label{fig:imp_NA}\includegraphics[width=0.45\textwidth]{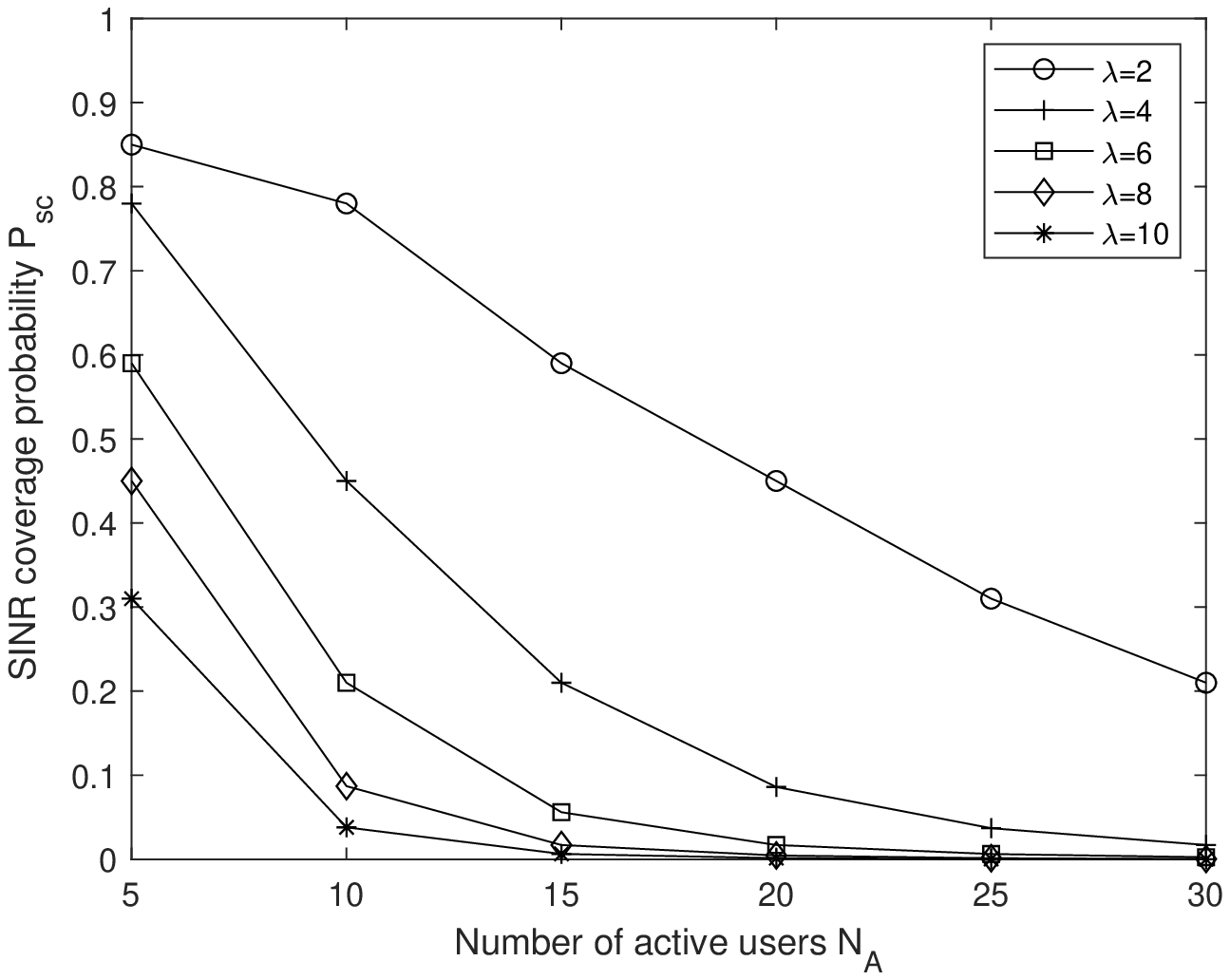}}
\caption{The impact of the average number of users $\nAVGpack$ and number of active users $\NAusers$ on the SINR coverage probability $\Psucc$ for a fixed number of time slots ($\nslot=20$) is presented in Figs. \ref{fig:model_vali} and \ref{fig:imp_NA}, respectively. In Fig. \ref{fig:model_vali}, the simulation and theoretical values are represented by dashed and solid lines, respectively. The gap in the results in Fig. \ref{fig:model_vali} can be mainly attributed to the lower number of $\NAusers$ and $\nAVGpack$. It can be seen that the simulation results become asymptotically equivalent to theoretical results at higher $\NAusers$ and $\nAVGpack$. Thus, only analytical results are presented in Fig. \ref{fig:imp_NA}. }\label{fig:model_beh}
\end{figure*}

\subsection{Model Validation and SINR Coverage Probability Behavior}
Fig. \ref{fig:model_vali} plots the SINR coverage probability $\Psucc$ within a transmission frame as a function of the average number of packets $\nAVGpack$ per active device, for a given number of active users $\NAusers$ at a given hovering location \textcolor{black}{while considering a maximum allowable latency of \textcolor{black}{$\frametime=10^{-3}$ s}}. The results are presented for $\NAusers=10,15,20$ active devices for a fixed number of time slots $(\nslot=20)$. We observe that both analytical and simulation results present the same trends, as $\NAusers$ and $\nAVGpack$ vary. The simulation results match reasonably well with the analytical results at severe traffic conditions, i.e., high $\NAusers$ and $\nAVGpack$ values. The gap between the simulation and the analytical results at light traffic conditions, i.e., low $\NAusers$ and $\nAVGpack$ values, can be attributed to assumptions made in the analytical framework. First, in the analytical framework, we estimate the number of singleton devices $\Nsing$ in a time slot using the collision free probability $\Pcf$ in \eqref{eq:Pcf} and the number of devices transmitting in each time slot. However, due to the probabilistic nature of this approach, $\Nsing$ is not always an integer. Thus, in the analytical framework, we consider the contribution of the fractional singleton devices to the SIC decoding process. Similarly, the interference at the decoding device also consists of fractional collided devices. In the simulation, no approximation on fractional devices is made and $\NAusers$ active devices within the frame randomly choose time slots to transmit their packets. Second, in the analytical framework, as stated in Assumption \ref{assum:order}, we assume distance-based ordering, instead of instantaneous received power-based ordering \cite{tabassum2017modeling}, at the UAV. This ignores the impact of small-scale fading and treats the closest device to the UAV as the device with the highest SINR. Thus, our assumption may not be true in severe fading environments. Nevertheless, as $\NAusers$ and $\nAVGpack$ increases, the simulation results quickly approach the analytical results. Thus, in the following, we only present numerical results using the proposed analytical framework, due to its reasonable accuracy.

Fig. \ref{fig:imp_NA} presents the impact of $\NAusers$ within the UAV serving zone on $\Psucc$ for different $\nAVGpack$ for a fixed number of time slots ($\nslot=20$) within a transmission frame with maximum latency of \textcolor{black}{$10^{-3}$ s}. We see that for fixed $\nAVGpack$ and $\nslot$, when $\NAusers$ increases, $\Psucc$ decreases. For a fixed number of time slots, when $\NAusers$ within a transmission frame increases, the number of devices transmitting at a given time slot increases. This decreases the MUSA collision probability $\Pcf$ and increases the number of undecodable MUSA sequence collided devices. On the other hand, due to an increasing number of transmitting devices in the given time slot, the number of singleton devices also increases. As the devices are distributed within a fixed serving area and transmit with equal power, an increasing number of singleton devices implies a decrease in power disparity within the SIC decodable devices. This decreases the performance of the SIC receiver and decreases the $\Psucc$. For a given $\NAusers$, when $\nAVGpack$ increases, $\Psucc$ decreases. When the number of packets generated by an active device increases, the probability that a given time slot is occupied by at least one device increases. This decreases $\Pcf$ and $\Nsing$ in the time slot, and decreases $\Psucc$ as explained above. Furthermore, if the number of packets generated by a device exceeds the fixed number of available time slots, the excess packets are not transmitted to the UAV at all. This further decreases $\Psucc$ at the UAV.

Overall, it can be seen from Fig. \ref{fig:imp_NA} that for a fixed number of time slots, as the number of active devices and/or the number of packets generated by each active device increases, the SINR coverage probability is severely degraded. This illustrates the fact that under an EoI, the system reliability is severely degraded unless the UAV changes its operating mode to accommodate the change in traffic conditions by changing the number of time slots, as stated in Remark \ref{rem:Psucc}.

\subsection{Impact of the Number of Time Slots}
In this subsection, we investigate the impact of the number of time slots on the SINR coverage probability and the short packet transmission error probability.

\begin{figure*}[t]
\centering
\subfigure[SINR coverage probability $\Psucc$ vs. number of time slots $\nslot$ for different $\nAVGpack=2,4,6,8,10$. ]{\label{fig:imp_ns}\includegraphics[width=0.45\textwidth]{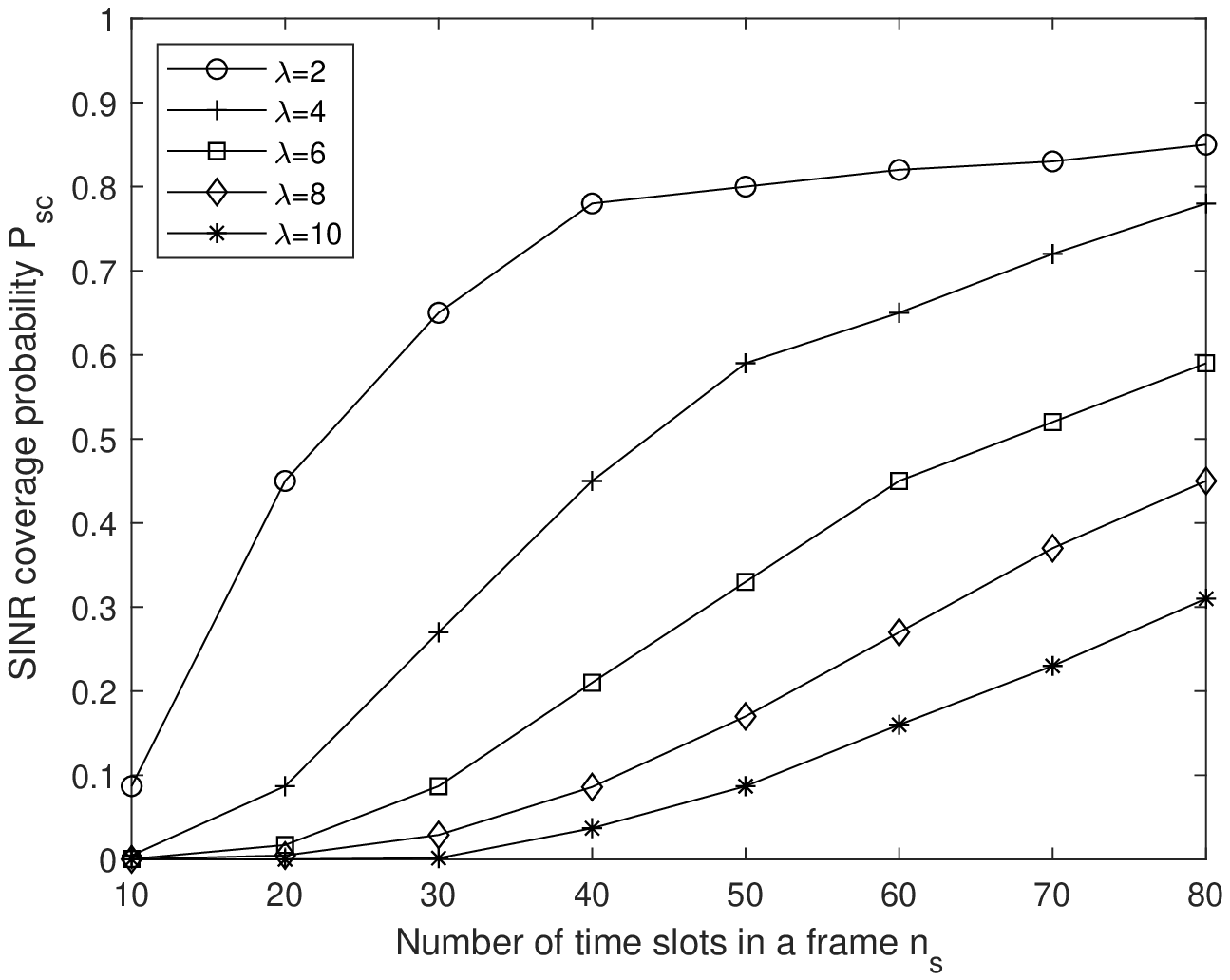}}
\subfigure[\textcolor{black}{Short packet transmission error probability $\epsilon$ vs. number of time slots $\nslot$ for different $\gamma=0,5,10$ dB.} ]{\label{fig:imp_ns_error}\includegraphics[width=0.45\textwidth]{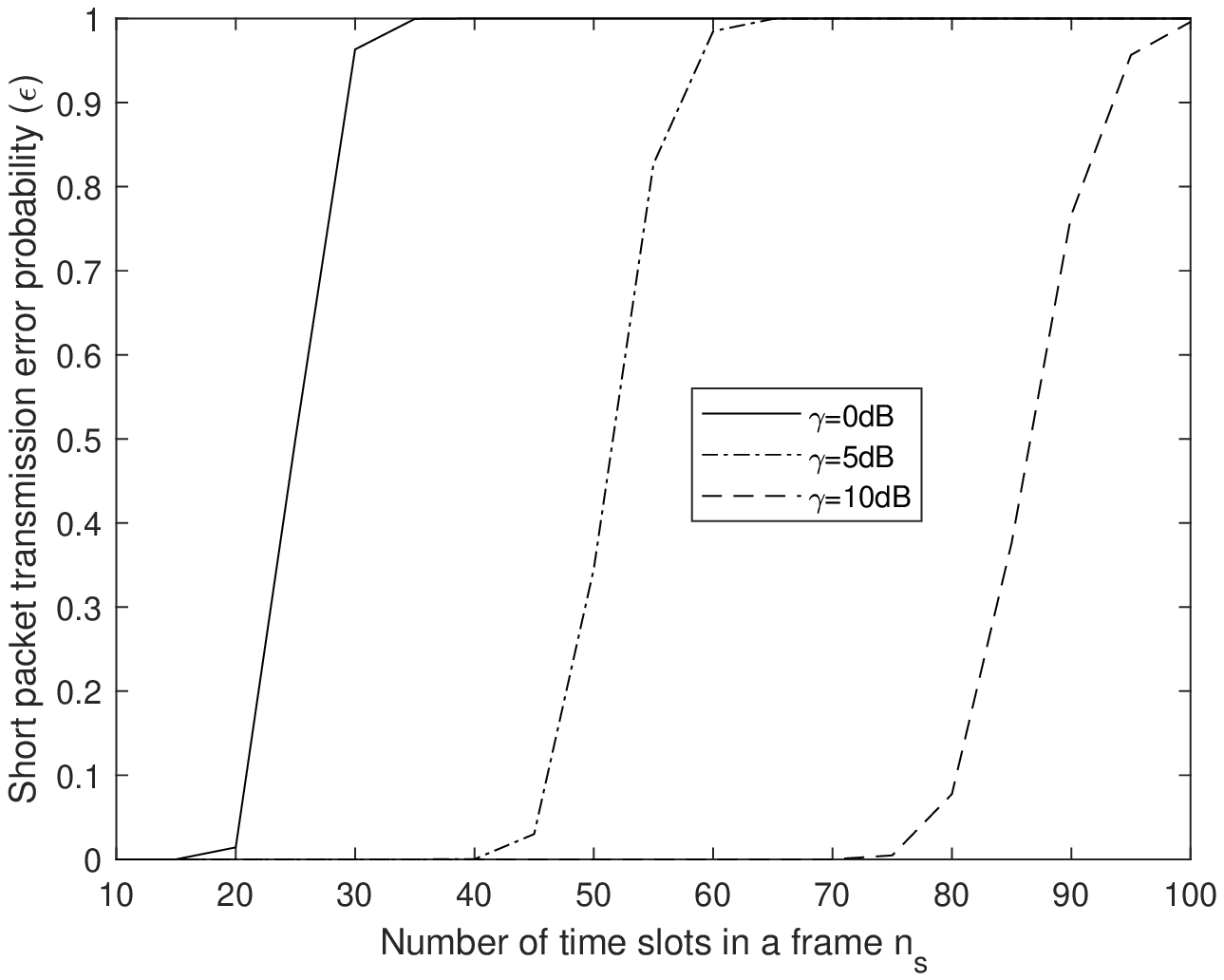}}
\caption{Impact of number of time slots on the SINR coverage probability and short packet transmission error probability.}\label{fig:ns}
\end{figure*}

Fig. \ref{fig:imp_ns} evaluates the impact on the number of time slots $\nslot$ for different $\nAVGpack$ for $\NAusers=20$ within a transmission frame with the maximum latency of \textcolor{black}{$10^{-3}$ s}. We see that for a given $\lambda$, when $\nslot$ increases, $\Psucc$ increases. Similar trend is observed for all the values of $\nAVGpack$. This is because when the number of time slots increases, the probability that a given time slot is occupied by more than one device decreases. As a result, the probability of MUSA collisions decreases and the number of singleton devices in a time slot decreases. When there are less singleton devices in a time slot, the performance of the SIC decoder increases. Thus, $\Psucc$ increases when $\nslot$ increases.

We also investigate the impact of number of time slots on short packet transmission errors for the IoT-UAV transmission. We present the impact of the number of time slots $\nslot$ on the short packet transmission error probability $\epsilon$ for different SINRs $\gamma$ in Fig. \ref{fig:imp_ns_error}. We can see that for a given $\nslot$, $\epsilon$ decreases when $\gamma$ increases. Similarly, for a given $\gamma$, $\epsilon$ increases when $\nslot$ increases, as stated in Remark \ref{rem:error}.

Overall we can see, from Figures \ref{fig:imp_ns} and \ref{fig:imp_ns_error} that there is a trade-off between $\Psucc$ and $\epsilon$. Therefore, both $\Psucc$ and $\epsilon$ must be considered simultaneously to optimize the number of time slots $\nslot$ while maximizing the performance of the IoT-UAV communication. Therefore, in the next subsection, we investigate the performance of the IoT-UAV transmission achieved by the adaptive transmission scheme proposed in Section \ref{sec:OP}.

\subsection{Proposed Scheme for Emergency Scenario}
We now present the performance of the proposed scheme in Section \ref{sec:OP} and compare with two benchmarks, namely, transmit power diversity scheme (TPDS) and a non-adaptive scheme (NAS). In the TPDS, power diversity is introduced to improve the SIC decoding by assigning random transmit powers to active devices within the power budget\footnote{One alternative to introduce power diversity is to allocate different power levels to the packets of the same device. However, this implies that some packets have higher priority over others. In our system model, we assume that all the packets are equally important. Another alternative is to vary the transmit power of devices while assuming that all the devices transmitting in a particular time slot have the knowledge of device locations, number of packets, and time slot allocations of other devices. However, this requires the devices to communicate with each other. Therefore, to compare the performance of our proposed scheme, we choose the TPDS as the most appropriate power diversity-based benchmark scheme.}. In the TPDS, $\nslot$ remains fixed; however, transmit power level of packets varies with the number of packets generated by a given device in a transmission frame. Suppose that the $i$th device has $L_{i}$ packets and each packet transmits with an equal power of $\frac{\powbud}{L_{i}}$. Since $L_{i}$ is a Poisson random variable, the transmit power of devices is random. In the NAS, neither the transmit power level nor the number of time slots changes with the number of packets generated by a user in a transmission frame.

\begin{figure}[t]
  \centering
  \includegraphics[width=0.5\textwidth]{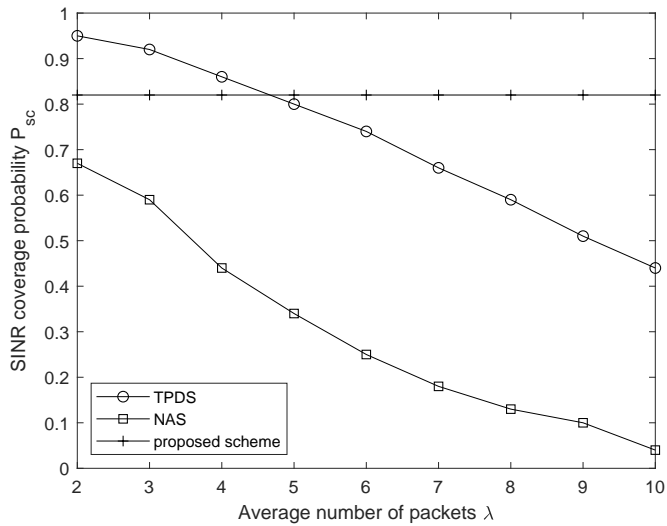}
  \caption{\textcolor{black}{Performance comparison of TPDS, NAS and proposed schemes with a maximum latency of $10^{-3}$ s for $\NAusers=10$.}}\label{fig:bench10}
\end{figure}

Fig. \ref{fig:bench10}  compares $\Psucc$ vs. $\nAVGpack$ for the proposed scheme, TPDS and NAS. In the proposed scheme, the maximum allowable latency and the short packet transmission error probability are constrained to \textcolor{black}{$10^{-3}$ s} and $10^{-5}$, respectively. The results are presented for $\NAusers=10$ active devices. In Fig. \ref{fig:bench10}, we see that both our scheme and the TPDS outperform the NAS. We then see that the TPDS outperforms the proposed scheme when $\nAVGpack$ is low, but then underperforms when $\nAVGpack$ increases. This is because that the TPDS randomizes the transmit power among devices by equally allocating $\Pmax$ among the packets within the device. At low $\nAVGpack$, the TPDS introduces enough power disparity between the transmitting devices in the time slot while maintaining a reasonable transmit power level. At high $\nAVGpack$, each packet of a given device is transmitted with a low transmit power, which decreases the SINR. Since the number of time slots remains unchanged in the TPDS, the MUSA collision probability increases and $\Nsing$ in a time slot increases, both dominating the benefit of power diversity in the TPDS. We further see that the performance of the proposed scheme remains stable regardless of increase in $\nAVGpack$, compared to the TPDS and NAS. Thus, we show that the proposed scheme has superior reliability and stability in comparison to the state-of-the-art TPDS at moderate to high average traffic rates. In the proposed scheme, all the devices transmit with fixed transmit power. According to Fig. \ref{fig:bench10}, it is evident, that incorporating power diversity to the proposed scheme would improve the performance of the proposed scheme for low average traffic. Thus, future work may consider maximizing the SINR coverage probability of active devices under stringent reliability and latency constraints in EoI by jointly optimizing the transmission slot duration and transmit power of active devices within the transmission frame.

\section{Conclusion}\label{sec:conclusion}
We proposed a novel adaptive transmission scheme for a UAV-assisted IoT-based environmental monitoring network where IoT devices transmit heterogeneous traffic under specific reliability and latency requirements. We employed MUSA-based GF-NOMA and short packet transmission to enable high-reliability and low-latency IoT-UAV transmission. To establish the proposed adaptive transmission scheme, we formulated an optimization problem to maximize the SINR coverage probability by optimizing the number of time slots in a transmission frame, subject to the short packet error constraint. Our results showed that, compared to benchmarks, the proposed scheme achieves superior reliability and stability at moderate to high average traffic rates while meeting the stringent low-latency constraints.

\begin{appendices}
\section{Proof of Lemma \ref{lemma:Lap_Is}}\label{app:laplace_Is}
The Laplace transform of $\Intsing$ can be written as
\begin{subequations}\label{eq:Is_proof_main}
\small
  \begin{align}
  \label{eq:Is_proof_a}
    \mathcal{L}_{\Intsing}(s)
    &=\mathbb{E}_{r,\hfad}\left[\exp\left(-s \sum_{i\in\PPPsing}\avgTransP\PLcoef (r_{i}^{2}+\hUAV^2)^{-\frac{\PLindex}{2}}\hfad_{i} \right)\right]\\ \nonumber
    &=\mathbb{E}_{\PPPsing}\left[\prod_{i\in\PPPsing}\mathbb{E}_{\hfad}
    \left[\exp(-s \avgTransP\PLcoef(r_{i}^{2}+\hUAV^2)^{-\frac{\PLindex}{2}}
    \hfad_{i})\right]\right]\\
    \label{eq:Is_proof_b}
    &=\mathbb{E}_{\PPPsing}\left[\prod_{i\in\PPPsing}\left(
    \frac{1}{1+s\avgTransP\PLcoef(r^{2}+\hUAV^2)^{-\frac{\PLindex}{2}}}
    \right)\right]\\
    \label{eq:Is_proof_c}
    &=\exp\left(-2 \pi\densing\int_{\hat{r}}^{\cellrad}\left(1-
  {\left(\frac{1}{1+s\avgTransP\PLcoef (r^{2}+\hUAV^2)^{-\frac{\PLindex}{2}}}\right)}\right)r dr\right).
  \end{align}
\end{subequations}
\normalsize
\noindent Here, \eqref{eq:Is_proof_a} is obtained by considering that the singleton devices in a given time slot follow a thinned PPP distribution $\PPPsing$ with intensity $\densing=\denori \Pcf$, where $\denori$ is the intensity of the original PPP distribution and $\Pcf$ is given in \eqref{eq:Pcf}. Moreover, \eqref{eq:Is_proof_b} is obtained by applying the moment generating function (MGF) of $\hfad$ which follows an exponential distribution with a unit mean. Furthermore, \eqref{eq:Is_proof_c} is obtained under the assumption that the collective interference experienced by the $k$th strongest device, which is located at a horizontal distance $\hat{r}$ from the UAV, is generated by the singleton devices located within the annulus with inner and outer radius of $\hat{r}$ and $\cellrad$. Therefore, \eqref{eq:lap_sing} is obtained by applying Campbell's theorem to \eqref{eq:Is_proof_b} and setting the lower and upper limits of integration of \eqref{eq:Is_proof_b} to $\hat{r}$ and $R$, respectively.

\vspace{-2mm}
\section{Proof of Theorem \ref{theo:OP}}\label{app:OP_sol}
The Lagrangian function for \eqref{eq:OP_main} is given by
\begin{subequations}\label{eq:lagrang}
\small
\begin{align*}
  \mathfrak{L}(\nslot,\Delta_i)&=-\Psucc
  +\Delta_1(\nslot-\NAusers\nAVGpack+\delta(\nAVGpack))
  +\Delta_2(\nAVGpack-\nslot)
  +\Delta_3 \\ \nonumber
  &\left(\mathbb{Q}\left[\sqrt{\frac{\bandwidth \frametime}{V\nslot}}
  \left(\log_{2}{(1\!+\!\gamma)-\frac{\databits\nslot}{\bandwidth \frametime}}
  \right)\right]-\errorProb\right)
  +\Delta_4 \\ \nonumber
  &\times(\nAVGpack_{\mathrm{min}}-\nAVGpack)
  +\Delta_5(\nAVGpack-\nAVGpack_{\mathrm{max}})
  +\Delta_6(r_{\mathrm{min}}-\cellrad),
  \end{align*}
\end{subequations}
\normalsize
\noindent where $\Delta_i, \forall i=1,\cdots,6$, corresponds to the Lagrangian multiplier associated with the $C_{i}$th constraint in \eqref{eq:OP_main}. The KKT conditions for \eqref{eq:OP_main} are written as
\begin{subequations}\label{eq:KKT_cond}
  \begin{align} \label{eq:kkt_c1}
    & \nslot-(\NAusers\nAVGpack+\delta(\nAVGpack)) \leq 0, \\ \label{eq:kkt_c2}
    & \nAVGpack-\nslot \leq 0,\\ \label{eq:kkt_c3}
    & \mathbb{Q}\left[\sqrt{\frac{\bandwidth \frametime}{V\nslot}}
  \left(\log_{2}{(1\!+\!\gamma)-\frac{\databits\nslot}{\bandwidth \frametime}}
  \right)\right]-\errorProb \leq 0,\\ \label{eq:kkt_c4}
    & \lambda_{\mathrm{min}}-\nAVGpack \leq 0,
    \nAVGpack-\nAVGpack_{\mathrm{max}} \leq 0,
    R_{\mathrm{min}}-\cellrad \leq 0,\\ \label{eq:dual_feas}
    & \Delta_1 \geq 0, \Delta_2 \geq 0, \Delta_3 \geq 0,
    \Delta_4 \geq 0, \Delta_5 \geq 0, \Delta_6 \geq 0,\\ \label{eq:comp_sl1}
    & \Delta_1(\nslot-\NAusers\nAVGpack+\delta(\nAVGpack))=0,\\ \label{eq:comp_sl2}
    & \Delta_2(\nAVGpack-\nslot)=0,\\ \label{eq:comp_sl3}
    & \Delta_3\left(\mathbb{Q}\left[\sqrt{\frac{\bandwidth \frametime}{V\nslot}}
  \left(\log_{2}{(1\!+\!\gamma)-\frac{\databits\nslot}{\bandwidth \frametime}}
  \right)\right]-\errorProb\right)=0,\\ \label{eq:comp_sl4}
  \small
    &\Delta_4(\nAVGpack_{\mathrm{min}}-\nAVGpack)=0,
    \Delta_5(\nAVGpack-\nAVGpack_{\mathrm{max}})=0,
    \Delta_6(r_{\mathrm{min}}-\cellrad)=0,\\ \label{eq:stat_cond}
    \normalsize
    & \frac{\partial}{d\partial}(\mathfrak{L}(\nslot,\Delta_i))=0,
  \end{align}
\end{subequations}

\normalsize
\noindent where \eqref{eq:kkt_c1}, \eqref{eq:kkt_c2}, \eqref{eq:kkt_c3}, and \eqref{eq:kkt_c4} correspond to the primal feasibility conditions associated with constraints $C_1-C_6$ in \eqref{eq:OP_main}, \eqref{eq:dual_feas} corresponds to the dual feasibility conditions associated with the Lagrangian multipliers in \eqref{eq:lagrang}, \eqref{eq:comp_sl1}, \eqref{eq:comp_sl2}, \eqref{eq:comp_sl3}, and \eqref{eq:comp_sl4} are the complementary slackness conditions applied to the constraints in \eqref{eq:OP_main} and \eqref{eq:stat_cond} is the stationary condition. The first derivative of the Lagrangian associated with aforementioned stationary condition is given by

\ifCLASSOPTIONdraftclsnofoot \vspace{-2mm}\fi
\small
\begin{align}\nonumber
  \frac{\partial}{\partial\nslot}(\mathfrak{L}(\nslot,\Delta_i))&=
  -\frac{\partial}{\partial \nslot}(\Psucc)+\Delta_1-\Delta_2
  +\Delta_3 \\ \label{eq:lag_1stderi}
  &\times\left(-\frac{\databits\log(2)}{2\sqrt{\bandwidth\frametime}}
  \nslot^{-\frac{1}{2}}+\frac{\log(1+\gamma)\sqrt{\bandwidth\frametime}}{2}
  \nslot^{-\frac{3}{2}}\right),
\end{align}

\begin{figure*}[t]
\begin{align}\nonumber
  &\frac{\partial}{\partial\nslot}{\Psucc}=\frac{1}{\NAusers\nAVGpack}\Bigg(
  \sum_{j=1}^{\Nsing}\Bigg(
  \prod_{k=1}^{j}\frac{3^{\Nsing-k}}{4^{\Nsing-1}}\frac{\Nsing !}{(\Nsing-k)!(k-1)!}
  \frac{1}{\barnes[1+j]}
  \imgI^{j(3+j)}3^{-\frac{1}{2}j(1+j-2\Nsing)}4^{(1-\Nsing)j}
  \\ \nonumber
  &\times\Lambda\Bigg(1-\frac{1}{4}j\nslot\pi\cellrad^2\tau
  \left(4\left(\frac{\threshold}{1+\threshold}
  \right)(1-y-x z)+3\left(\frac{\kappa \threshold}{1+\kappa\threshold}\right)\right)
  \Bigg)
  +\Lambda\nslot\barnes[1-\Nsing]^{-1+j}\barnes[1+j-\Nsing]\\
  \label{eq:Psucc_deri}
  &\times(\barnes[2-\Nsing]\Gamma(\Nsing))^{-j}
  \left(j\left(1+\frac{1}{\Nsing}-2\coth^{-1}(7)\right)\right)
  +(-j+\Nsing)(\harmonic_{j-\Nsing}-\harmonic_{-\Nsing})\NAusers\tau(y-z x)
  \Bigg)
  \Bigg).
\end{align}
\rule{18.2cm}{0.5pt}
\end{figure*}

\normalsize
\noindent where $\frac{\partial}{\partial\nslot}\Psucc$ is given by \eqref{eq:Psucc_deri} at the top of this page, with $x=\Plambda$, $y=\Pcf$, $\kappa=\left(\frac{\hUAV^2+\frac{\cellrad^2}{4}}{\hUAV^2+\frac{9\cellrad^2}{16}}
\right)^{\frac{\PLindex}{2}}$, and $\barnes[\cdot]$, $\harmonic_{[\cdot]}$, and $\daleth$ denoting the Barnes-G function, Harmonic function, and the Euler's constant, respectively. $\Lambda$ and $\tau$ are given by
\small
\begin{align}\nonumber
  \Lambda &=\exp\Bigg(\frac{1}{4}j\Bigg(-\frac{4\left(\hUAV^2+\frac{\cellrad^2}{4}\right)^
  \frac{\PLindex}{2}\threshold\noiseP}{\avgTransP\PLcoef}-4\NAslot x(1-y)\\ \label{eq:Lambda}
  &\times\left(\frac{\threshold}{1+\threshold}\right)\dencol-3\NAusers x y
  \left(\frac{\kappa\threshold}{1+\kappa\threshold}\right)
  \Bigg)\Bigg)
\end{align}
\normalsize
and
\begin{equation}\label{eq:tau}
\small
  \tau=-\frac{1}{\nslot^2}e^{-\nAVGpack}\sum_{L=1}^{\nslot-1}\frac{L\nAVGpack^{L}}{L!}
  +e^{-\nAVGpack}\sum_{L=\nslot}^{\nAVGpack+L_{lim}}\frac{\nAVGpack^{L}}{\Gamma(1+L)}
  \left(\daleth-\harmonic_{L}+\log(\nAVGpack)\right),
\end{equation}

\normalsize
\noindent respectively. From the complementary slackness condition, we know that either $\Delta_i$ is zero or the associated constraint function is zero for any given $i$. We consider one of the possible cases $\Delta_1$ is not slack, i.e., $\Delta_1$ exists and $\Delta_2,\Delta_3$ do not exist. Note that we ignore $\Delta_4,\Delta_5,\Delta_6$ as the corresponding constraints are not functions of $\nslot$. Thus, substituting $\Delta_1\neq0$, and $\Delta_2$ ,$\Delta_3=0$ to the stationary condition, we obtain
\begin{equation}\label{eq:delta1_case}
  -\frac{\partial}{\partial \nslot}\Psucc+\Delta_1=0.
\end{equation}

Considering the complimentary slackness condition \eqref{eq:comp_sl1}, we obtain $\tilde{n}_{s}=\NAusers\nAVGpack+\delta(\nAVGpack)$. Substituting $\tilde{n}_{s}$ to \eqref{eq:delta1_case}, we derive $\Delta_1$ as
\begin{equation}\label{eq:delta1}
  \Delta_1=\frac{\partial}{\partial \nslot (\nslot=\tilde{n}_{s})}\Psucc.
\end{equation}

It can be shown that the dual feasibility condition associated with $\Delta_1$ is satisfied as $\Delta_1\geq 0$. Also, $\tilde{n}_{s}$ satisfy the primal feasibility conditions. Therefore, $\tilde{n}_{s}$ and $\Delta_1$ satisfy all the KKT conditions and it can be shown that $\tilde{n}_{s}$ is the optimal solution to the $\mathbf{OP}$ in \eqref{eq:OP_main} when all the constraints in $\mathbf{OP}$ are slack, except for the first constraint. Thus, the optimal solution to the $\mathbf{OP}$ is given as \textcolor{black}{$\nslot^{\ast}=\min{(\NAusers\nAVGpack+\delta(\nAVGpack)
,n_{\errorProb})}$} where $n_{\errorProb}$ is the solution to $\mathbb{Q}\left[\sqrt{\frac{\bandwidth \frametime}{n_\errorProb}} \left(\log{(1+\gamma)-\frac{\databits}{\bandwidth \frametime}}\log(2) n_\errorProb\right)\right]-\errorProb=0$. \textcolor{black}{Therefore, $\tilde{\nslot}=\lfloor\nslot^{\ast}\rfloor$}. Note that this is the only solution to the $\mathbf{OP}$ in \eqref{eq:OP_main} as all the other possible cases of $\Delta_{i}$ violate one or more KKT conditions. 

\end{appendices}

 \bibliographystyle{IEEEtran}

\end{document}